\pdfoutput=1
\documentclass[reprint,aps,prx,showpacs,twocolumn]{revtex4-1}

\usepackage{natbib}
\usepackage{graphicx}
\usepackage[utf8]{inputenc}
\usepackage{mathptmx}
\usepackage{epsfig,amsopn}
\usepackage{graphicx}
\usepackage{amsmath,amssymb}
\usepackage{natbib,color}
\usepackage{braket}
\usepackage{float}
\usepackage{enumerate}
\usepackage[normalem]{ulem}
\usepackage{todonotes}
\allowdisplaybreaks[1]
\usepackage{threeparttable}

\def\bea{\begin{eqnarray}}
\def\eea{\end{eqnarray}}

\begin{document}

\title{Growth-laws and invariants from ribosome biogenesis in lower Eukarya}

\author{Sarah Kostinski$^{1}$}
\email{skostinski@tauex.tau.ac.il}

\author{Shlomi Reuveni$^{1,2}$}
\email{shlomire@tauex.tau.ac.il}

\affiliation{\noindent \textit{$^{1}$School of Chemistry, The Center for Physics and Chemistry of Living Systems, \& The Mark Ratner Institute for Single Molecule Chemistry, Tel Aviv University, Tel Aviv 6997801, Israel}}

\affiliation{\noindent \textit{$^{2}$The Raymond and Beverly Sackler Center for Computational Molecular and Materials Science, Tel Aviv University, Tel Aviv 6997801, Israel}}

\date{\today}

\begin{abstract} 
\noindent

\noindent 

Eukarya and Bacteria are the most evolutionarily distant domains of life, which is reflected by differences in their cellular structure and physiology.  For example, Eukarya feature membrane-bound organelles such as nuclei and mitochondria, whereas Bacteria have none.  The greater complexity of Eukarya renders them difficult to study from both an experimental and theoretical perspective.  However, encouraged by a recent experimental result showing that budding yeast (a unicellular eukaryote) obeys the same proportionality between ribosomal proteome fractions and cellular growth rates as Bacteria, we derive a set of relations describing eukaryotic growth from first principles of ribosome biogenesis.  We recover the observed ribosomal protein proportionality, and then continue to obtain two growth-laws for the number of RNA polymerases synthesizing ribosomal RNA per ribosome in the cell. These growth-laws, in turn, reveal two invariants of eukaryotic growth, i.e. quantities predicted to be conserved by Eukarya regardless of growth conditions. The invariants, which are the first of their kind for Eukarya, clarify the coordination of transcription and translation kinetics as required by ribosome biogenesis, and link these kinetic parameters to cellular physiology.  We demonstrate application of the relations to the yeast \textit{S. cerevisiae} and find the predictions to be in good agreement with currently available data.  We then outline methods to quantitatively deduce several unknown kinetic and physiological parameters.  The analysis is not specific to \textit{S. cerevisiae} and can be extended to other lower (unicellular) Eukarya when data become available. The relations may also have relevance to certain cancer cells which, like bacteria and yeast, exhibit rapid cell proliferation and ribosome biogenesis.  

\vspace{5mm}

\end{abstract}

\maketitle

\noindent

\section{Introduction}

Recent advances in biological physics have led to the discovery of quantitative relations, or ``laws,'' describing bacterial growth~\cite{Klumpp2008, ScottHwaReview, KlumppPNAS, Pugatch, Alon1, Alon2}, gene expression~\cite{Salman,AmirNatComm}, and cell size control~\cite{AmirPRL,Taheri-Araghi}; see Ref.~\cite{JunReview} for a review and historical perspective.  Whether similar relations apply to Eukarya, a domain of life that is evolutionarily distant from Bacteria, remains unclear.  Eukarya and Bacteria differ in many ways; with respect to cellular organization, Eukarya contain membrane-bound organelles such as nuclei and mitochondria that Bacteria lack altogether.  This spatial partitioning of the eukaryotic cell affects numerous processes involving the transport of essential macromolecules.  One such process is the generation of new ribosomes, termed ribosome biogenesis, where ribosomes are the central macromolecular machines of protein synthesis in the cell.  In Eukarya, ribosome biogenesis requires that ribosomal subunits be transported from the nucleolus to the nucleoplasm, and eventually to the cytoplasm via nuclear pores, while simultaneously undergoing maturation~\cite{Woolford, Henras, Thomson}.  The eukaryotic ribosome is also substantially larger than its bacterial counterpart, having 25 proteins which have no equivalent in bacterial ribosomes~\cite{BenShem}.  Furthermore, in contrast to a few non-essential assembly factors in Bacteria, ribosome assembly in yeast, a unicellular eukaryote, requires about 200 accessory proteins which do not even form part of the mature ribosome.  If just one accessory protein is missing, ribosome biogenesis cannot proceed~\cite{Karbstein,Dinman}.    

In light of these additional complexities, it is not surprising that quantitative relations for eukaryotic growth are still lacking.  An indication that it might be possible to generalize certain bacterial growth-laws to lower (unicellular) Eukarya was reported in 2017 by Metzl-Raz et al.~\cite{Barkai}.  There the authors demonstrated that ribosomal proteome fractions in budding yeast are proportional to cellular growth rates, as previously observed for Bacteria~\cite{Scott2010,Scott}.  Underlying this proportionality is the coupling between cell growth and ribosome biogenesis~\cite{Dai}, i.e. that cell doubling requires a commensurate doubling of ribosomes.  The latter leads to an autocatalytic loop and a fundamental bound on cellular growth rates since ribosomal proteins (r-proteins) can only be made by other ribosomes~\cite{Bionum, Dill, REP}.   

The important discovery made in Ref.~\cite{Barkai} supports the notion that ribosome biogenesis is growth-limiting in Eukarya just as in Bacteria.  Cytoplasmic ribosomes are not only composed of r-protein though; in fact, their main constituent is ribosomal RNA (rRNA).  In Bacteria, rRNA is produced by RNA polymerases (RNAPs), which in turn are made by ribosomes.  We recently showed that this process leads to another bound on cellular growth rates and to growth-laws which were verified for the bacterium \textit{E. coli}~\cite{KR,Speed-Limit}.  But rRNA production in Eukarya diverges from that in Bacteria: Bacteria have just one kind of RNAP in the cell while Eukarya have at least three, two of which --  RNA polymerase I (RNAP I) and RNA polymerase III (RNAP III) -- are involved in the production of rRNA.  Moreover, the coordination mechanisms of rRNA and r-protein production in Eukarya are entirely different from Bacteria~\cite{NomuraThoughts}.  Yet, despite the greater complexity, here we show that simple growth-laws can still be established for Eukarya. 

In this work, we provide a model of ribosome biogenesis in lower Eukarya and study its implications for cell physiology and growth.  The analysis yields two growth-laws and two invariants which are the first of their kind for Eukarya.  We then corroborate the relations using currently available data for the model organism \textit{S. cerevisiae} (budding yeast), and discuss additional data that will be needed for full verification of all the relations.  The growth-laws and invariants offer quantitative predictions and provide a theoretical framework for future studies on \textit{S. cerevisiae} and similar organisms.  Our work suggests that the ribosome composition in \textit{S. cerevisiae} is optimized for cell growth as in \textit{E. coli}, but more data are required for verification.

The remainder of this paper is structured as follows. In Section~\ref{II}, we mathematically formulate the kinetics of ribosome production in lower Eukarya.  From these equations we obtain upper bounds on the cellular growth rate, which are given in Section~\ref{III}.  We show that the bounds are uniquely maximized for a specific ribosome composition in Section~\ref{IV}.  Growth rate maximization yields three distinct growth-laws, which are derived in Section~\ref{V}. These include the already known proportionality between r-protein fractions and growth rates, as well as two additional growth-laws for RNAP I and RNAP III which make rRNA in eukaryotic cells. The growth-laws, in turn, yield two invariants. These conserved quantities, which illustrate the coordination of rRNA and r-protein production in the cell, are discussed in Section~\ref{VI}. In Section~\ref{VII}, we provide a physical interpretation of the growth-laws and invariants in terms of proteome fractions.  Section~\ref{VIII} offers a case study of the model organism \textit{S. cerevisiae}, showing that predictions from the growth-laws are consistent with currently available data.  The analysis offers quantitative predictions for, e.g., the number of ribosomes in the cell, the number of RNAPs I and RNAPs III required for rRNA production, and the coupling between the rates of translation, transcription, and cell growth.  Finally, we discuss application of the invariants to determine activities of RNAPs I and III once their proteome fractions are known to better accuracy.  Section~\ref{conc} concludes this work with a discussion on future research directions.  More detailed derivations, and data for the case study of \textit{S. cerevisiae}, are provided in the Appendices. 

\section{Kinetics of ribosome biogenesis} 
\label{II}

Ribosomes are critical to cellular growth since they produce all protein in the cell, where protein comprises the largest fraction ($\sim \! 40$\%) of the cell's dry mass~\cite{Bionum} (BNID: 104157). These protein-producing machines are ubiquitous: a rapidly growing yeast cell contains more than 200,000 ribosomes~\cite{Warner, WaldronLacroute75}. Ribosomes, in turn, are composed of ribosomal protein (r-protein) and ribosomal RNA (rRNA), which account for large fractions of the cell's proteome and RNA content.  For example, in \textit{S.~cerevisiae}, r-protein and rRNA are estimated to comprise up to a third of the proteome mass~\cite{Barkai} and $\sim \! 80$\% of the total RNA mass~\cite{Warner}, respectively. To better understand the kinetics of ribosome biogenesis, we proceed to 
write a set of differential equations describing the average production rates of r-protein and rRNA in the cell.  

\begin{figure}[t!]
\begin{centering}
\includegraphics[width=0.5 \textwidth]{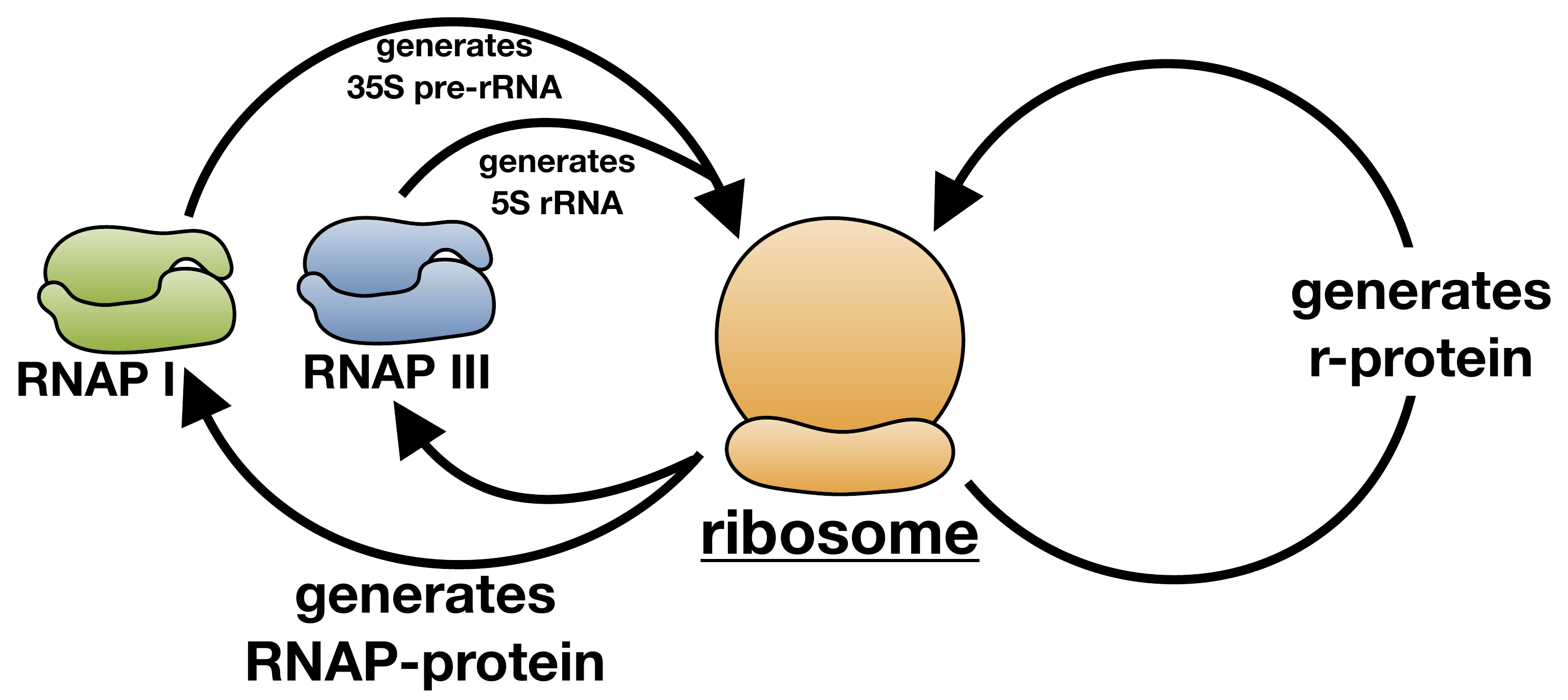}
\end{centering}
\vspace{-5mm}
\caption{\textbf{Ribosome biogenesis in lower Eukarya.} Ribosomal protein (r-protein) is synthesized directly by ribosomes, as illustrated by the autocatalytic loop on the right. Meanwhile ribosomal RNA (rRNA) is synthesized by RNA polymerases I and III, symbolized by the top left arrows.  In lower Eukarya,  RNA polymerase I generates the 35S precursor rRNA which later yields the mature 25S, 18S, and 5.8S rRNAs, while RNA polymerase III generates the 5S rRNA.  RNA polymerases, in turn, are made of protein that is synthesized by ribosomes (bottom left arrows). Here we show that these autocatalytic processes give rise to multiple growth-laws and invariants that characterize central aspects of eukaryotic cell growth.}
\label{yeastautocatalytic}
\end{figure}

Ribosomes make r-protein directly (Fig.~\ref{yeastautocatalytic}). The r-protein production rate, measured in amino acids per unit time, can be written as
\begin{equation}
    \frac{d(\text{r-protein})}{dt} = k_{\text{ribo}} \, \phi^{\text{r-prot}}_{\text{ribo}} \, f^{\text{active}}_{\text{ribo}} \cdot N_{\text{ribo}},
\label{rproteindiffeq}
\end{equation}
where $N_{\text{ribo}}$ is the number of ribosomes in the cell, $f^{\text{active}}_{\text{ribo}}$ is the fraction of ribosomes which are active, $\phi^{\text{r-prot}}_{\text{ribo}}$ is the fraction of active ribosomes making specifically r-protein, and $k_{\text{ribo}}$ is the average peptide elongation rate of an active ribosome. During exponential growth, there is little to no protein degradation~\cite{MiloBook,ProteinDegradation}, and so Eq.~(\ref{rproteindiffeq}) describes the accumulation rate of r-protein in the cell.  Note that the fraction of active ribosomes making specifically r-protein, $\phi^{\text{r-prot}}_{\text{ribo}}$, is equivalent to the time fraction an active ribosome spends synthesizing r-protein.  These two interpretations are based on either an ensemble or time average: The latter entails tracking the time an active ribosome spends on r-protein synthesis, where time fractions are obtained by averaging over long times, i.e. spanning many cell generations.  In the ensemble picture, the fraction of active ribosomes engaged in r-protein synthesis is instead estimated from snapshots of the cell taken at arbitrary times.   

\begin{figure*}[t!]
\begin{centering}
\includegraphics[width=0.75\textwidth]{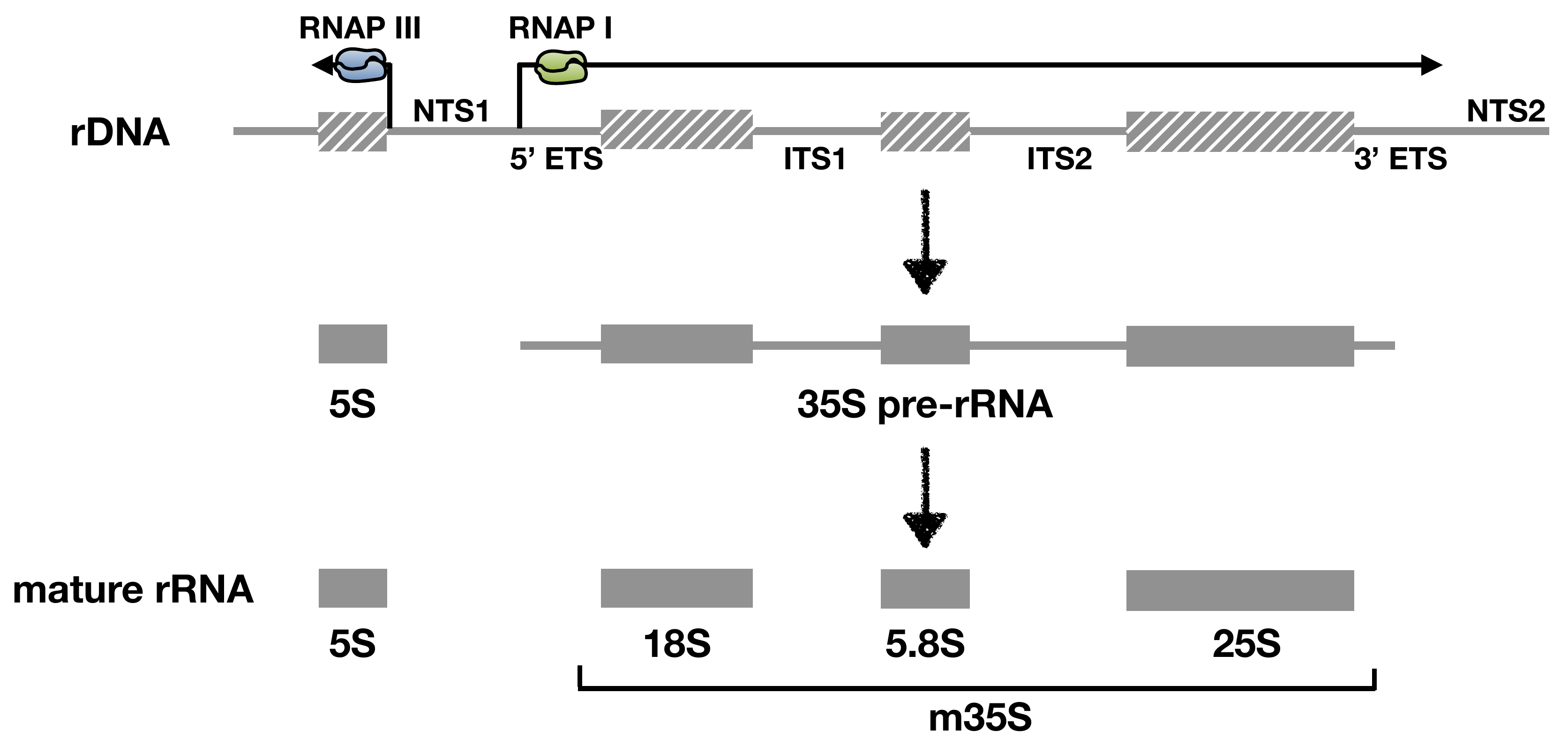}
\end{centering}
\vspace{-2mm}
\caption{\textbf{Transcription and processing of rRNA.} RNAP I (green) transcribes the 35S pre-rRNA from the ribosomal DNA (rDNA) locus, while the RNAP III (blue) separately transcribes the 5S rRNA (121 nucleotides long).  In yeast, the 35S pre-rRNA contains a total of 1504 spacer nucleotides from: ITS1 (361 nt), ITS2 (232 nt), 5'ETS (700 nt), and 3'ETS (211 nt)~\cite{SGD}, where ITS and ETS denote an internally transcribed spacer and an externally transcribed spacer, respectively.  These spacer nucleotides are later processed away to yield the mature 35S-derived rRNAs (collectively abbreviated as m35S): 18S (1800 nt), 5.8S (158 nt), and 25S (3396 nt)~\cite{Woolford}.}
\label{rRNAprocessing}
\end{figure*}

To make rRNA, which is the main constituent of cytoplasmic ribosomes, eukaryotic cells use two types of RNAPs: RNAP I and RNAP III (Fig.~\ref{yeastautocatalytic}, Fig.~\ref{rRNAprocessing}).  RNAPs themselves are composed solely of protein, and so equations for the accumulation rates of RNAP I- and RNAP III-protein can be written similarly to the above: 
\begin{eqnarray}
    \frac{d(\text{RPI-protein})}{dt} = k_{\text{ribo}} \, \phi^{\text{RPI}}_{\text{ribo}} \, f^{\text{active}}_{\text{ribo}} \cdot N_{\text{ribo}} \label{RNAPIdiffeq} \, , \\ 
    \frac{d(\text{RPIII-protein})}{dt} = k_{\text{ribo}} \, \phi^{\text{RPIII}}_{\text{ribo}} \, f^{\text{active}}_{\text{ribo}} \cdot N_{\text{ribo}} \, , \label{RNAPIIIdiffeq}
\end{eqnarray}
where in Eq.~(\ref{RNAPIdiffeq}) we have used the fraction $\phi^{\text{RPI}}_{\text{ribo}}$ of active ribosomes dedicated to the synthesis of RNAP I-protein, and in Eq.~(\ref{RNAPIIIdiffeq}) the fraction $\phi^{\text{RPIII}}_{\text{ribo}}$ of active ribosomes dedicated to RNAP III-protein synthesis.

The production rates of the mature 18S, 25S, and 5.8S rRNAs, which are generated by RNAP I, and of the 5S rRNA generated by RNAP III, can be expressed in nucleotides per unit time as:
\begin{eqnarray}
    \frac{d(\text{m35S})}{dt} = k_{\text{RPI}} \, \phi^{\text{m35S}}_{\text{RPI}} \, f^{\text{active}}_{\text{RPI}} \cdot N_{\text{RPI}} \, , \label{35Sdiffeq} \\ 
    \frac{d(\text{5S})}{dt} = k_{\text{RPIII}} \, \phi^{\text{5S}}_{\text{RPIII}} \, f^{\text{active}}_{\text{RPIII}} \cdot N_{\text{RPIII}}  \, , \label{5Sdiffeq}
\end{eqnarray}
where m35S on the left-hand side of Eq.~(\ref{35Sdiffeq}) represents the number of nucleotides in all mature 35S-derived rRNAs in the cell, i.e. the 18S, 25S and 5.8S rRNAs, while 5S in Eq.~(\ref{5Sdiffeq}) denotes the number of nucleotides in all 5S rRNAs in the cell (Fig.~\ref{rRNAprocessing}). In addition, 
$N_{\text{RPI}}$ and $N_{\text{RPIII}}$ represent the number of RNA polymerases I and III, respectively; the fraction of RNAPs I that are active is given by $f^{\text{active}}_{\text{RPI}}$, and $f^{\text{active}}_{\text{RPIII}}$ is the active fraction of RNAPs III.  The quantity $\phi^{\text{m35S}}_{\text{RPI}}$ denotes the fraction of active RNAPs I making 18S, 25S, and 5.8S rRNAs (spacer nucleotides not included), while $\phi^{\text{5S}}_{\text{RPIII}}$ is the fraction of active RNAPs III making 5S rRNA. Finally, $k_{\text{RPI}}$ and $k_{\text{RPIII}}$ are the average rRNA chain elongation (transcription) rates of an active RNAP I and an active RNAP III, respectively.

\section{Upper bounds on cellular growth rate}
\label{III}

The number of ribosomes $N_{\text{ribo}}$ in the cell can be approximated as
\begin{eqnarray}
    N_{\text{ribo}} \simeq \frac{\text{r-protein}}{N^{\text{a.a.}}_{\text{ribo}}}, 
    \label{Nriboapprox1}
\end{eqnarray}
where r-protein is measured in units of the number of amino acids per cell, while $N_{\text{ribo}}^{\text{a.a.}}$ is the number of amino acids in the ribosome. Similarly, one can estimate $N_{\text{ribo}}$ as
\begin{eqnarray}
   N_{\text{ribo}} \simeq \frac{\text{m35S}}{N^{\text{nucl}}_{\text{m35S}}} \simeq \frac{\text{5S}}{N^{\text{nucl}}_{\text{5S}}} \, , \label{Nriboapprox2} 
\end{eqnarray}
where $N^{\text{nucl}}_{\text{m35S}}$ is the combined total number of nucleotides in the 18S, 25S, 5.8S rRNAs, while $N^{\text{nucl}}_{\text{5S}}$ is the number of nucleotides per 5S rRNA.  It is important to note that $N^{\text{nucl}}_{\text{m35S}}$ does not include flanking or spacer nucleotides in the 35S-precursor rRNA, as they are cleaved away during rRNA processing (see Fig.~\ref{rRNAprocessing}). Eqs.~(\ref{Nriboapprox1}) and (\ref{Nriboapprox2}) provide overestimates of $N_{\text{ribo}}$ since all r-protein and rRNA in the cell is assumed to be fully assembled into  ribosomes.  This can be compensated for, however, by the ribosomal activity $f^{\text{active}}_{\text{ribo}}$, which accounts for nascent r-protein and rRNA as inactive.

Similarly, we approximate the number of RNAPs I and III as   
\begin{eqnarray}
    N_{\text{RPI}} \simeq \frac{\text{RPI-protein}}{N^{\text{a.a.}}_{\text{RPI}}} \, , \label{RNAPIapprox}\\
    N_{\text{RPIII}} \simeq \frac{\text{RPIII-protein}}{N^{\text{a.a.}}_{\text{RPIII}}} \, ,
    \label{RNAPIIIapprox}
\end{eqnarray}
where the numerators are the number of amino acids in RNAP I-protein and RNAP III-protein in the cell, respectively.  Meanwhile the denominators $N^{\text{a.a.}}_{\text{RPI}}$ and $N^{\text{a.a.}}_{\text{RPIII}}$ are the number of amino acids in each RNAP I and each RNAP III, respectively.  Again, note that Eq.~(\ref{RNAPIapprox}) and Eq.~(\ref{RNAPIIIapprox}) provide overestimates of $N_{\text{RPI}}$ and $N_{\text{RPIII}}$.

Equations~(\ref{rproteindiffeq})--(\ref{5Sdiffeq}) can be simplified using the approximations of Eqs.~(\ref{Nriboapprox1})--(\ref{RNAPIIIapprox}) to give upper bounds on protein and rRNA production rates in the cell.  Substituting Eq.~(\ref{Nriboapprox1}) into Eq.~(\ref{rproteindiffeq}) then yields
\begin{equation}
    \frac{d(\text{r-protein})}{dt} = k_{\text{ribo}} \, \phi^{\text{r-prot}}_{\text{ribo}} \, f^{\text{active}}_{\text{ribo}} \cdot \frac{\text{r-protein}}{N^{\text{a.a.}}_{\text{ribo}}} \, ,
\end{equation}
whose solution is exponential for balanced growth, in which the parameters $k_{\text{ribo}}$, $\phi^{\text{r-prot}}_{\text{ribo}}$, $f^{\text{active}}_{\text{ribo}}$ are constant by definition.  The cellular growth rate $\mu$ is then bounded by
\begin{equation}
    \mu \equiv \frac{\ln(2)}{T_d} \leq \frac{k_{\text{ribo}} \, \phi^{\text{r-prot}}_{\text{ribo}} \, f^{\text{active}}_{\text{ribo}}}{N^{\text{a.a.}}_{\text{ribo}}} \, ,
    \label{bound1}
\end{equation}
where $T_{d}$ is the cellular doubling time.  In the bacterium \textit{E. coli}, it was shown that Eq.~(\ref{bound1}) is not only a bound but in fact an approximate equality~\cite{KR}.  The resulting proportionality between the r-protein proteome fraction $\phi^{\text{r-prot}}_{\text{ribo}}$ and the cellular growth rate $\mu$ has been called a ``growth-law of ribosome synthesis''~\cite{JunReview, Scott2010}. In yeast, the same proportionality was established by Kief \& Warner~\cite{KiefWarner} and more comprehensively by Metzl-Raz, et al.~\cite{Barkai}, but the proportionality factor $k_{\text{ribo}}f^{\text{active}}_{\text{ribo}}/N^{\text{a.a.}}_{\text{ribo}}$ in Eq.~(\ref{bound1}) still requires direct experimental verification. Note that the latter need not be constant for $\phi^{\text{r-prot}}_{\text{ribo}} \propto \mu$ to hold. This important point, which is often overlooked, will be discussed further in Section~\ref{VIII}.

Two additional bounds on the cellular growth rate can be derived by making a similar substitution of Eq.~(\ref{Nriboapprox1}) in Eqs.~(\ref{RNAPIdiffeq}) and (\ref{RNAPIIIdiffeq}):
\begin{eqnarray}
    \frac{d(\text{RPI-protein})}{dt} = k_{\text{ribo}} \, \phi^{\text{RPI}}_{\text{ribo}} \, f^{\text{active}}_{\text{ribo}} \cdot \frac{\text{r-protein}}{N^{\text{a.a.}}_{\text{ribo}}} \, , \label{RNAPIsub} \\ 
    \frac{d(\text{RPIII-protein})}{dt} = k_{\text{ribo}} \, \phi^{\text{RPIII}}_{\text{ribo}} \, f^{\text{active}}_{\text{ribo}} \cdot \frac{\text{r-protein}}{N^{\text{a.a.}}_{\text{ribo}}}, \label{RNAPIIIsub}
\end{eqnarray}
\noindent and using the approximations of Eqs.~(\ref{RNAPIapprox})--(\ref{RNAPIIIapprox}) in Eqs.~(\ref{35Sdiffeq})--(\ref{5Sdiffeq}):
\begin{eqnarray}
    \frac{d(\text{m35S})}{dt} = k_{\text{RPI}} \, \phi^{\text{m35S}}_{\text{RPI}} \, f^{\text{active}}_{\text{RPI}} \cdot \frac{\text{RPI-protein}}{N^{\text{a.a.}}_{\text{RPI}}} \, , \label{35SdiffeqwRNAPapprox} \\ 
    \frac{d(\text{5S})}{dt} = k_{\text{RPIII}} \, \phi^{\text{5S}}_{\text{RPIII}} \, f^{\text{active}}_{\text{RPIII}} \cdot \frac{\text{RPIII-protein}}{N^{\text{a.a.}}_{\text{RPIII}}}. \label{5SdiffeqwRNAPapprox}
\end{eqnarray}
Taking a time derivative of Eqs.~(\ref{35SdiffeqwRNAPapprox})--(\ref{5SdiffeqwRNAPapprox}) and using Eqs.~(\ref{RNAPIsub})--(\ref{RNAPIIIsub}) for the production rates of RNAP I- and RNAP III-protein, together with the approximations of Eqs.~(\ref{Nriboapprox1})--(\ref{Nriboapprox2}), then yields 
\begin{align}
    & \frac{d^{2}(\text{m35S})}{dt^{2}} = \frac{k_{\text{RPI}} \, \phi^{\text{m35S}}_{\text{RPI}} \, f^{\text{active}}_{\text{RPI}}}{ N^{\text{a.a.}}_{\text{RPI}}} \cdot \frac{k_{\text{ribo}} \, \phi^{\text{RPI}}_{\text{ribo}} \, f^{\text{active}}_{\text{ribo}}}{N^{\text{nucl}}_{\text{m35S}}} \cdot \text{m35S} \, , \label{35Sorder2diffeq}\\ 
    & \frac{d^{2}(\text{5S})}{dt^{2}} = \frac{k_{\text{RPIII}} \, \phi^{\text{5S}}_{\text{RPIII}} \, f^{\text{active}}_{\text{RPIII}}}{N^{\text{a.a.}}_{\text{RPIII}}} \cdot \frac{k_{\text{ribo}} \, \phi^{\text{RPIII}}_{\text{ribo}} \, f^{\text{active}}_{\text{ribo}}}{N^{\text{nucl}}_{\text{5S}}} \cdot \text{5S}. \label{5Sorder2diffeq}
\end{align}
The exponential solutions of Eqs.~(\ref{35Sorder2diffeq})--(\ref{5Sorder2diffeq}) reveal that the cellular growth rate $\mu$ is bounded by
\begin{equation}
    \mu \leq \sqrt{\frac{k_{\text{RPI}} \, \phi^{\text{m35S}}_{\text{RPI}} \, f^{\text{active}}_{\text{RPI}}}{ N^{\text{a.a.}}_{\text{RPI}}} \cdot \frac{k_{\text{ribo}} \, \phi^{\text{RPI}}_{\text{ribo}} \, f^{\text{active}}_{\text{ribo}}}{N^{\text{nucl}}_{\text{m35S}}}} \, ,
    \label{bound2}
\end{equation}
and
\begin{equation}
    \mu \leq \sqrt{\frac{k_{\text{RPIII}} \, \phi^{\text{5S}}_{\text{RPIII}} \, f^{\text{active}}_{\text{RPIII}}}{N^{\text{a.a.}}_{\text{RPIII}}} \cdot \frac{k_{\text{ribo}} \, \phi^{\text{RPIII}}_{\text{ribo}} \, f^{\text{active}}_{\text{ribo}}}{N^{\text{nucl}}_{\text{5S}}}} \, .
    \label{bound3}
\end{equation}
In contrast to bacteria, which have just one type of RNA polymerase and thus one bound on cellular growth rate originating from the production of rRNA~\cite{KR}, lower Eukarya must satisfy two bounds -- one originating from each type of RNA polymerase producing rRNA.

\begin{figure*}[t!]
\begin{centering}
\includegraphics[width=0.75\textwidth]{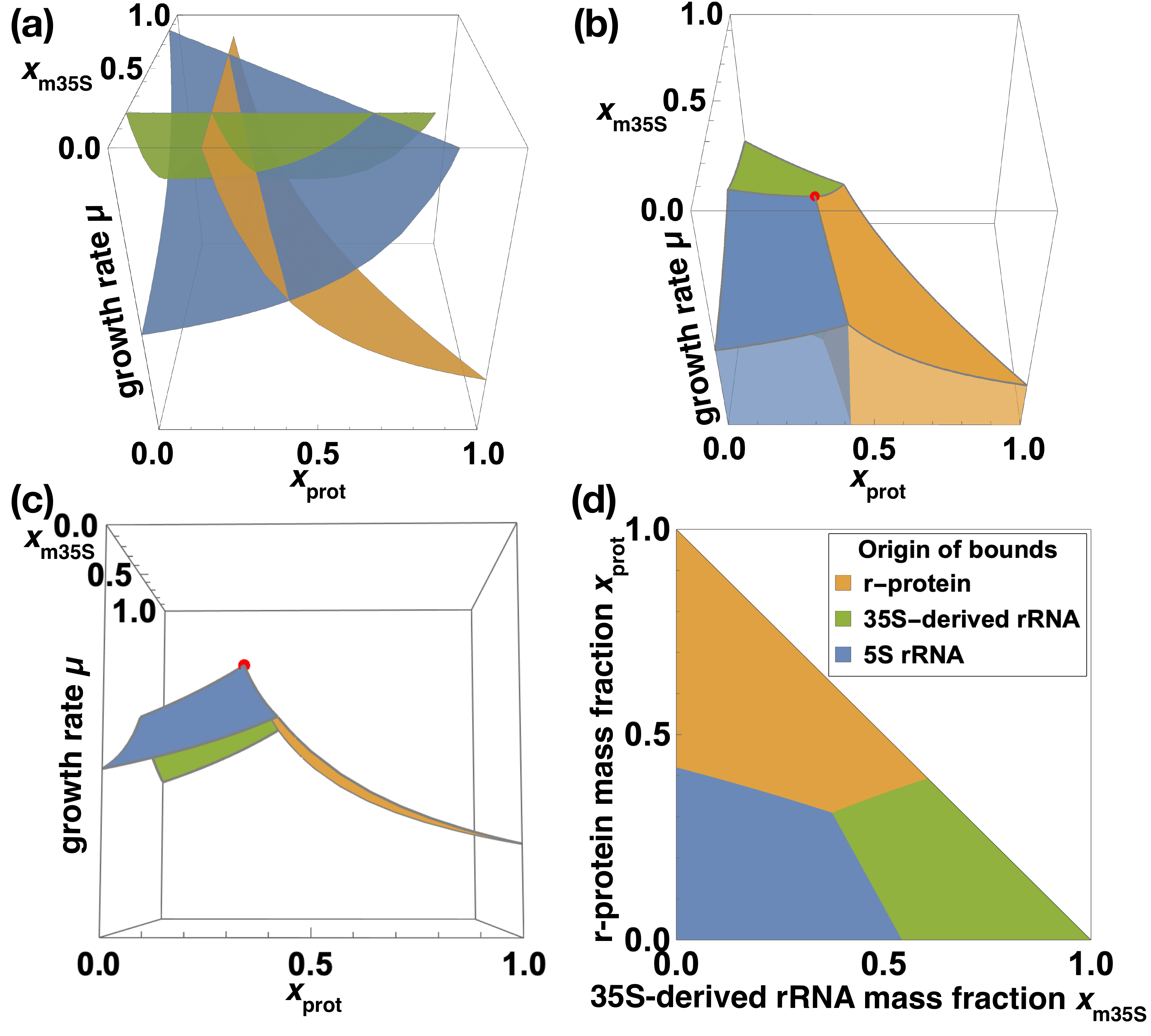}
\end{centering}
\vspace{-2mm}
\caption{\textbf{Bounds on cellular growth rate from the production of r-protein (orange), 35S-derived rRNA (green), and 5S rRNA (blue). (a)} We plot the three bounds [Eqs.~(\ref{bound1xy}), (\ref{bound2xy}), (\ref{bound3xy})] versus $x_{\text{prot}}$, the mass fraction of the ribosome which is r-protein [Eq.~(\ref{xdef})], and versus $x_{\text{m35S}}$, the mass fraction of the ribosome which is mature 35S-derived rRNA  [Eq.~(\ref{ydef})]. Each surface in the $x_{\text{prot}}$-$x_{\text{m35S}}$-$\mu$ space corresponds to one of the three bounds.  They are made partially transparent so that their intersections are visible.  \textbf{(b)}  Plotted is the minimum of all surfaces for every possible $x_{\text{prot}}$ and $x_{\text{m35S}}$.  Accessible growth rates lie in the shaded regions below the surfaces.  Each shaded region is colored according to its most restricting bound.  \textbf{(c)} Side view of the minimal surfaces which intersect to form a cusp, marked by the red point.  This point corresponds to the maximal cellular growth rate permitted by the three bounds.  \textbf{(d)} The point where all three bounds intersect is clearly seen when viewed from below the bounds.  For visual clarity we chose the intersection point to lie further in the interior of the $x_{\text{prot}}$-$x_{\text{m35S}}$ plane; in reality it will lie closer to the diagonal line ($x_{\text{m35S}}=1-x_{\text{prot}}$) since the mass fraction of 5S rRNA, $1-x_{\text{prot}}-x_{\text{m35S}}$, is small.  The parameter values used to generate this figure are available in Appendix~\ref{B}.}
\label{surfaces}
\end{figure*}

\section{Graphical representation of bounds}\label{IV}

The derived bounds can be expressed in terms of two variables describing ribosome composition: $x_{\text{prot}}$, the mass fraction of the ribosome which is protein, and $x_{\text{m35S}}$, the mass fraction of the ribosome which is mature rRNA derived from the 35S precursor (18S, 25S, 5.8S rRNAs).  Defining the ribosome mass as $M_{\text{ribo}} \simeq N^{\text{a.a.}}_{\text{ribo}} \, m_{\text{a.a.}} + N^{\text{nucl}}_{\text{m35S}} \, m_{\text{nucl}} + N^{\text{nucl}}_{\text{5S}} \, m_{\text{nucl}}$, where $m_{\text{a.a.}}$ and $m_{\text{nucl}}$ are the average masses of an amino acid and nucleotide in the cell, respectively, gives \begin{equation}
    x_{\text{prot}} = \frac{N^{\text{a.a.}}_{\text{ribo}} \, m_{\text{a.a.}}}{M_{\text{ribo}}} \, .
    \label{xdef}
\end{equation}
The bound of Eq.~(\ref{bound1}) then becomes 
\begin{equation}
    \mu \leq \frac{m_{\text{a.a.}}}{M_{\text{ribo}}} \cdot \frac{k_{\text{ribo}} \, \phi_{\text{ribo}}^{\text{r-prot}} \, f^{\text{active}}_{\text{ribo}}}{x_{\text{prot}}} \, .
   \label{bound1xy}
\end{equation}
Furthermore, the mass fraction of the ribosome which is rRNA, $1-x_{\text{prot}}$, can be partitioned into 35S-derived and 5S rRNA masses.  Defining $x_{\text{m35S}}$ as the mass fraction of the former, i.e. the 18S, 25S, and 5.8S rRNAs, 
\begin{equation}
    x_{\text{m35S}} = \frac{N^{\text{nucl}}_{\text{m35S}} \, m_{\text{nucl}}}{M_{\text{ribo}}},
    \label{ydef}
\end{equation}
yields $N^{\text{nucl}}_{\text{m35S}} = \frac{M_{\text{ribo}}}{m_{\text{nucl}}} \cdot x_{\text{m35S}}$ and $N^{\text{nucl}}_{\text{5S}} = \frac{M_{\text{ribo}}}{m_{\text{nucl}}} \cdot (1-x_{\text{prot}}-x_{\text{m35S}})$.  Substituting these relations into the remaining two bounds of Eqs.~(\ref{bound2})--(\ref{bound3}) yields
\begin{equation}
    \mu \leq \sqrt{\frac{k_{\text{RPI}} \, \phi^{\text{m35S}}_{\text{RPI}} \, f^{\text{active}}_{\text{RPI}}}{ N^{\text{a.a.}}_{\text{RPI}}} \cdot \frac{m_{\text{nucl}}}{M_{\text{ribo}}} \cdot \frac{k_{\text{ribo}} \, \phi^{\text{RPI}}_{\text{ribo}} \, f^{\text{active}}_{\text{ribo}}}{x_{\text{m35S}}}} \, ,
    \label{bound2xy}
\end{equation}
\begin{equation}
    \mu \leq \sqrt{\frac{k_{\text{RPIII}} \, \phi^{\text{5S}}_{\text{RPIII}} \, f^{\text{active}}_{\text{RPIII}}}{N^{\text{a.a.}}_{\text{RPIII}}} \cdot \frac{m_{\text{nucl}}}{M_{\text{ribo}}} \cdot \frac{k_{\text{ribo}} \, \phi^{\text{RPIII}}_{\text{ribo}} \, f^{\text{active}}_{\text{ribo}}}{1-x_{\text{prot}}-x_{\text{m35S}}}} \, .
    \label{bound3xy}
\end{equation}
The functional forms of the three bounds are thus $\mu \leq a/x_{\text{prot}}$ [Eq.~(\ref{bound1xy})], $\mu \leq b/\sqrt{x_{\text{m35S}}}$ [Eq.~(\ref{bound2xy})], and $\mu \leq c/\sqrt{1-x_{\text{prot}}-x_{\text{m35S}}}$ [Eq.~(\ref{bound3xy})], where $a$, $b$, and $c$ are positive constants. 

A set of bounds is unique to its particular growth condition, as every growth condition specifies different values of the constants $\{a, b, c\}$.  For a given growth condition, each bound defines a surface in the three-dimensional $x_{\text{prot}}$-$x_{\text{m35S}}$-$\mu$ space, as illustrated in Fig.~\ref{surfaces}a.  The volume which lies under the union of these three surfaces represents cellular growth rates accessible to the organism, as a function of ribosome composition (Fig.~\ref{surfaces}b).  The maximal cellular growth rates mutually satisfying two bounds are defined by the line which intersects the two corresponding surfaces.  Hence there are three lines defined by the three possible pairs of bounds (Fig.~\ref{surfaces}c), which can be written parametrically in terms of one free variable, e.g. $x_{\text{prot}}$ or $x_{\text{m35S}}$ (Appendix \ref{A}). The point at which the three lines intersect, i.e. the cusp of the union of the three surfaces, is obtained at an optimal ribosome composition that defines the maximum possible cellular growth rate satisfying all three bounds (Fig.~\ref{surfaces}c).  This point, where all three surfaces meet, is clearly seen when viewed from below (Fig.~\ref{surfaces}d).

A set of bounds from ribosome biogenesis, similar to those derived above, was first obtained for bacteria~\cite{KR}. There, it was shown that the $1:2$ protein to RNA mass ratio in the \textit{E. coli} ribosome is optimal in that it offers the maximal growth rate permitted by the bounds in a variety of growth conditions.  A similar principle may also apply to Eukarya, but direct verification of this hypothesis is currently not possible due to missing data. Verification will require simultaneous measurements of the growth rate and all parameters in Eqs.~(\ref{bound1xy}), (\ref{bound2xy}), and (\ref{bound3xy}), in various growth conditions.  However, even for a well-studied model organism like \textit{S. cerevisiae}, such a dataset is currently unavailable. 

The situation described above calls for comprehensive measurements of biologically relevant kinetic and physiological parameters in the yeast \textit{S. cerevisiae} and in other Eukarya,  similar to those done for \textit{E. coli}~\cite{BremerDennis}. While collecting such data is expected to be challenging and time-consuming, it will significantly advance our understanding of yeast, and more generally, of Eukarya.  In the case of \textit{E. coli}, a comprehensive dataset was key in recognizing that the bacterium achieves the maximal growth rate permitted by ribosome biogenesis.  This finding led to a previously unrecognized growth-law and an invariant of bacterial growth~\cite{KR}.  In lieu of complete datasets for Eukarya like \textit{S. cerevisiae}, we posit that their bounds can also be considered as approximate equalities, just as in Bacteria.  It yields a number of insights: In the following sections, we derive growth-laws and invariants for Eukarya, showing that the resulting predictions are in good agreement with currently available data. These results self-consistently support the postulate of growth rate maximization, and shed new light on the coordination of transcription and translation kinetics as required by ribosome biogenesis. It also allows one to deduce numerical values of unknown kinetic and physiological parameters in the yeast \textit{S. cerevisiae}.  

\section{Growth-laws from growth rate maximization}
\label{V}

In analogy to the bacterial case, we interpret the upper bounds of Eq.~(\ref{bound1}) and Eqs.~(\ref{bound2})--(\ref{bound3}) as approximate equalities.  Eq.~(\ref{bound1}) then simplifies to
\begin{equation}
    \boxed{ \tau_{\text{r-prot}} \cdot \mu \simeq f^{\text{active}}_{\text{ribo}} \, \phi^{\text{r-prot}}_{\text{ribo}}}  \, ,
    \label{growthlaw1}
\end{equation}
where on the left-hand side we have defined $\tau_{\text{r-prot}} = N^{\text{a.a.}}_{\text{ribo}} / k_{\text{ribo}} $ as the average time it takes a ribosome to synthesize a full set of r-proteins. 

Two other relations, or ``growth-laws,'' result from the three bounds found earlier, assuming that cells achieve the optimal growth rate.  For example, squaring Eqs.~(\ref{bound1}) and (\ref{bound2}) and setting their right-hand sides equal, while keeping one power of $\mu$ from Eq.~(\ref{bound1}), yields 
\begin{equation}
        \boxed{\tau_{\text{m35S}} \cdot \mu  \simeq \frac{(\phi^{\text{m35S}}_{\text{RPI}} \, f^{\text{active}}_{\text{RPI}} \phi^{\text{RPI}}_{\text{ribo}} ) \, / N^{\text{a.a.}}_{\text{RPI}}  }{\phi^{\text{r-prot}}_{\text{ribo}} / N^{\text{a.a.}}_{\text{ribo}}}} \, ,
        \label{growthlaw2}
\end{equation}
where we have defined $\tau_{\text{m35S}} = N^{\text{nucl}}_{\text{m35S}} / k_{\text{RPI}}$ as the average time it takes an RNAP I to synthesize a set of 18S, 25S, 5.8S rRNAs.  Similarly, defining $\tau_{\text{5S}} = N^{\text{nucl}}_{\text{5S}} / k_{\text{RPIII}} $ as the average time for an RNAP III to synthesize a 5S rRNA, from Eqs.~(\ref{bound1}) and (\ref{bound3}) we obtain
\begin{equation}
  \boxed{\tau_{\text{5S}} \cdot \mu \simeq \frac{ (\phi^{\text{5S}}_{\text{RPIII}} \, f^{\text{active}}_{\text{RPIII}} \, \phi^{\text{RPIII}}_{\text{ribo}}) / N^{\text{a.a.}}_{\text{RPIII}} }{\phi^{\text{r-prot}}_{\text{ribo}} / N^{\text{a.a.}}_{\text{ribo}}}} \, .
  \label{growthlaw3}
\end{equation}
A simple physical interpretation of the growth-laws in Eqs.~(\ref{growthlaw2}) and (\ref{growthlaw3}) will be discussed in a subsequent section.

\section{Invariants of cellular growth}
\label{VI}

To eliminate the explicit dependence on growth rate in the relations derived above, we divide the first growth-law [Eq.~(\ref{growthlaw1})] by the second [Eq.~(\ref{growthlaw2})], and multiply the numerator and denominator of the left-hand side by $m_{\text{nucl}} \, m_{\text{a.a.}}$.  Recognizing that $ m_{\text{a.a.}} N^{\text{a.a.}}_{\text{ribo}} = x_{\text{prot}} \, M_{\text{ribo}}$ and $m_{\text{nucl}} N^{\text{nucl}}_{\text{m35S}} = x_{\text{m35S}} \, M_{\text{ribo}}$, we obtain:
\begin{equation}
    \boxed{\frac{x_{\text{prot}}}{x_{\text{m35S}}} \simeq \frac{m_{\text{a.a.}}}{m_{\text{nucl}}} \cdot \frac{N^{\text{a.a.}}_{\text{RPI}}}{N^{\text{a.a.}}_{\text{ribo}}} \cdot \frac{k_{\text{ribo}} }{k_{\text{RPI}}} \cdot \frac{\phi^{\text{r-prot}}_{\text{ribo}} f^{\text{active}}_{\text{ribo}} \phi^{\text{r-prot}}_{\text{ribo}}}{\phi^{\text{m35S}}_{\text{RPI}} \, f^{\text{active}}_{\text{RPI}} \, \phi^{\text{RPI}}_{\text{ribo}}} } \, .
    \label{inv1}
\end{equation}
On non-evolutionary timescales, the ribosome composition is fixed and hence the ratio between the r-protein and 35S-derived rRNA mass fractions on the left-hand side of Eq.~(\ref{inv1}) is constant.  The translation and transcription parameters on the right-hand side must therefore be coordinated so as to satisfy this constraint.  That is, the numerical values of these parameters may vary between growth conditions, but in such a way that the right-hand side of the equation remains constant and equal to the left.  The right-hand side of Eq.~(\ref{inv1}) can thus be viewed as non-trivial invariant of eukaryotic growth, i.e. it is predicted to remain constant irrespective of growth conditions.

Similarly, dividing Eqs.~(\ref{growthlaw2}) by (\ref{growthlaw3}) and multiplying numerator and denominator by $m_{\text{nucl}}$, immediately reveals an invariant quantity via the mass ratio between 35S-derived rRNA and 5S rRNA:
\begin{equation}
    \boxed{\frac{x_{\text{m35S}}}{1-x_{\text{prot}}-x_{\text{m35S}}} \simeq \frac{N^{\text{a.a.}}_{\text{RPIII}}}{N^{\text{a.a.}}_{\text{RPI}}} \cdot \frac{k_{\text{RPI}}}{k_{\text{RPIII}}} \cdot \frac{\phi^{\text{m35S}}_{\text{RPI}} \, f^{\text{active}}_{\text{RPI}} \, \phi^{\text{RPI}}_{\text{ribo}}}{\phi^{\text{5S}}_{\text{RPIII}} \, f^{\text{active}}_{\text{RPIII}} \, \phi^{\text{RPIII}}_{\text{ribo}} } } \, .
    \label{inv2}
\end{equation}
There are many ways of expressing the two independent invariants. For example, an equivalent to Eq.~(\ref{inv1}) is obtained upon division of Eq.~(\ref{growthlaw1}) by Eq.~(\ref{growthlaw3}) to yield $x_{\text{prot}}/(1-x_{\text{prot}}-x_{\text{m35S}})$, i.e. the ratio between the r-protein and the 5S rRNA mass fractions (Appendix~\ref{C}).  Physical interpretations of the invariants are discussed in the following section. 

\section{Interpretation of growth-laws and invariants using proteome fractions}
\label{VII}

In the case that the parameters $\phi^{\text{r-prot}}_{\text{ribo}}$, $\phi^{\text{RPI}}_{\text{ribo}}$, and $\phi^{\text{RPIII}}_{\text{ribo}}$ cannot be measured directly, they can be approximated by proteome fractions.  In the absence of active and differential degradation among proteins, the fraction $\phi^{\chi}_{\text{ribo}}$ of active ribosomes making a protein of type $\chi$ is equal to the proteome fraction of $\chi$, i.e. ($\chi$-protein)/(total protein), since all protein in the cell is synthesized by ribosomes at an average rate $k_{\text{ribo}}$.  In \textit{E. coli} this approximation holds well because active protein degradation is negligible~\cite{MiloBook}.  In Eukarya, active degradation may be more significant in, e.g., stressful conditions and hence the proteome fraction approximation should be used only when suitable.  

A recent study on turnover rates of 3,160 proteins in exponentially growing \textit{S. cerevisiae} revealed a median protein half-life of 2.18 hr, which matches the corresponding cellular doubling time (2.0 $\pm$ 0.1 hr)~\cite{ProteinDegradation}. Differential protein degradation was also measured. Specifically, the median half-life of ribosomal proteins (1.7 hr) was found to be $\sim\!20$\% lower than the overall protein half-life; however, the authors of the study note that this difference may be an artifact of the measurement method.  Moreover, while some proteins in yeast seem to be actively degraded even in exponential growth, nearly all proteins exhibit half-lives close to the cellular doubling time.  It was thus concluded that active degradation of protein in exponential growth is small, and that the replacement rate of the proteome is dominated by growth and division.  Similar protein turnover trends were observed in human cells~\cite{Boisvert,Gawron}.  Thus it appears that in exponential growth, the quantities $\phi^{\text{r-prot}}_{\text{ribo}}$, $\phi^{\text{RPI}}_{\text{ribo}}$, and $\phi^{\text{RPIII}}_{\text{ribo}}$ can be approximated by their respective proteome fractions. 

The growth-laws of Eqs.~(\ref{growthlaw1}), (\ref{growthlaw2}), and (\ref{growthlaw3}) become more transparent with the proteome fraction approximations. In the first growth-law [Eq.~(\ref{growthlaw1})]: $\tau_{\text{r-prot}} \cdot \mu = f^{\text{active}}_{\text{ribo}} \, \phi^{\text{r-prot}}_{\text{ribo}}$, the quantity $\phi^{\text{r-prot}}_{\text{ribo}}$ can be interpreted as the proteome fraction of r-protein in the cell.  The right-hand side can then be interpreted as the active r-protein proteome fraction.  Alternatively, since $f^{\text{active}}_{\text{ribo}}$ is the fraction of ribosomes that are active, of which a fraction $\phi^{\text{r-prot}}_{\text{ribo}}$ is synthesizing r-protein, their product $f^{\text{active}}_{\text{ribo}} \, \phi^{\text{r-prot}}_{\text{ribo}}$ is the fraction of all ribosomes in the cell that are active and synthesizing r-protein.  Thus the first growth-law in Eq.~(\ref{growthlaw1}) becomes:
\begin{equation}
    \tau_{\text{r-prot}} \cdot \mu \simeq \text{fraction of ribosomes making r-protein} \, .
    \label{growthlaw1interpretation}
\end{equation}

In the case of the second growth-law [Eq.~(\ref{growthlaw2})], the product $\phi^{\text{m35S}}_{\text{RPI}} \, f^{\text{active}}_{\text{RPI}} \phi^{\text{RPI}}_{\text{ribo}}$ in the numerator can be interpreted as the proteome fraction of RNAP I-protein actively synthesizing 18S, 25S, and 5.8S rRNAs.  In the denominator, $\phi^{\text{r-prot}}_{\text{ribo}}$ is the r-protein proteome fraction.  Multiplying the right-hand side of Eq.~(\ref{growthlaw2}) by the quantity (total protein)/(total protein) then reveals it to be a ratio between the number of RNAPs I making mature rRNA and the number of ribosomes in the cell:
\begin{equation}
    \tau_{\text{m35S}} \cdot \mu \simeq  \frac{\text{\# of RNAPs I making m35S (18S/25S/5.8S)}}{\text{\# of ribosomes}}  \, .
    \label{growthlaw2interpretation}
\end{equation}
Note that Eq.~(\ref{growthlaw2}) can also be interpreted for the number of active RNAPs I, all of which are dedicated to the synthesis of the 35S precursor rRNA (flanking and spacer nucleotides included, see Fig. \ref{rRNAprocessing}).  Because $N^{\text{nucl}}_{\text{35S}}/N^{\text{nucl}}_{\text{m35S}}=\phi^{\text{35S}}_{\text{RPI}} / \phi^{\text{m35S}}_{\text{RPI}}$, where $N^{\text{nucl}}_{\text{35S}}$ is the number of nucleotides in the 35S pre-rRNA and $\phi^{\text{35S}}_{\text{RPI}}$ denotes the fraction of active RNAPs I (all making the 35S pre-rRNA), we obtain
\begin{equation}
     \tau_{\text{35S}} \cdot \mu \simeq  \frac{\text{\# of active RNAPs I}}{\text{\# of ribosomes}} \, ,
    \label{growthlaw2interpretation2}
\end{equation}
where $\tau_{\text{35S}}=N^{\text{nucl}}_{\text{35S}}/k_{\text{RPI}}$. An analogous interpretation can be made for the third growth-law [Eq.~(\ref{growthlaw3})], yielding
\begin{equation}
    \tau_{\text{5S}} \cdot \mu \simeq \frac{\text{\# of RNAPs III making 5S}}{\text{\# of ribosomes}}  \, .
    \label{growthlaw3interpretation}
\end{equation}

The invariants of Eqs.~(\ref{inv1}) and (\ref{inv2}) can also be interpreted using proteome fractions.  The first
[Eq.~(\ref{inv1})] simplifies to (Appendix~\ref{D}) 
\begin{equation}
    \frac{x_{\text{prot}}}{x_{\text{m35S}}}\simeq \frac{m_{\text{a.a.}}}{m_{\text{nucl}}} \cdot \frac{k_{\text{ribo}} }{k_{\text{RPI}}}  \cdot \frac{\text{\# of ribosomes making r-protein}}{\text{\# of RNAPs I making 18S/25S/5.8S}}  \, .
\label{inv1int}
\end{equation}
Meanwhile the second invariant contains a similar ratio:
\begin{equation}
    \frac{x_{\text{m35S}}}{1-x_{\text{prot}}-x_{\text{m35S}}} \simeq \frac{k_{\text{RPI}}}{k_{\text{RPIII}}} \cdot \frac{\text{\# of RNAPs I making 18S/25S/5.8S}}{\text{\# of RNAPs III making 5S}} \, , 
\label{inv2int}
\end{equation}
which demonstrates that RNAPs I and RNAPs III are coordinated for the stoichiometric production of rRNA.  Similar to the case of \textit{E. coli}~\cite{KR}, we expect the quantities on the right-hand sides of Eqs.~(\ref{inv1int})--(\ref{inv2int}) to be invariant for an exponentially growing eukaryote, regardless of external conditions. The numerical values of these invariants are set by the left-hand sides of the equations, and may thus differ from organism to organism in accordance with the endogenous ribosome composition.

\section{\textit{S. cerevisiae} as a case study}
\label{VIII}

To demonstrate the predictive potential of the relations derived above, we apply them to the model organism \textit{S. cerevisiae} using currently available data.  There has not yet been a systematic study of an eukaryote in a specific growth condition which includes all relevant parameters, as was done for the bacterium \textit{E. coli}~\cite{BremerDennis}.  However, we collected typical parameter ranges from various sources to serve as benchmark values (Appendix~\ref{E}) and were able to recover a number of results.  For example, below we deduce the dependence of ``ribosomal efficiency'' ($k_{\text{ribo}} \, f^{\text{active}}_{\text{ribo}}$) on growth rate, the number of RNAPs I per RNAP III required for rRNA production, and the number of ribosomes in the cell.  These results encourage future experiments to verify the remaining predictions.  We also outline methods to infer the activities of RNAP I and RNAP III once more data become available.

\begin{figure*}[t!]
\begin{centering}
\includegraphics[width=0.75\textwidth]{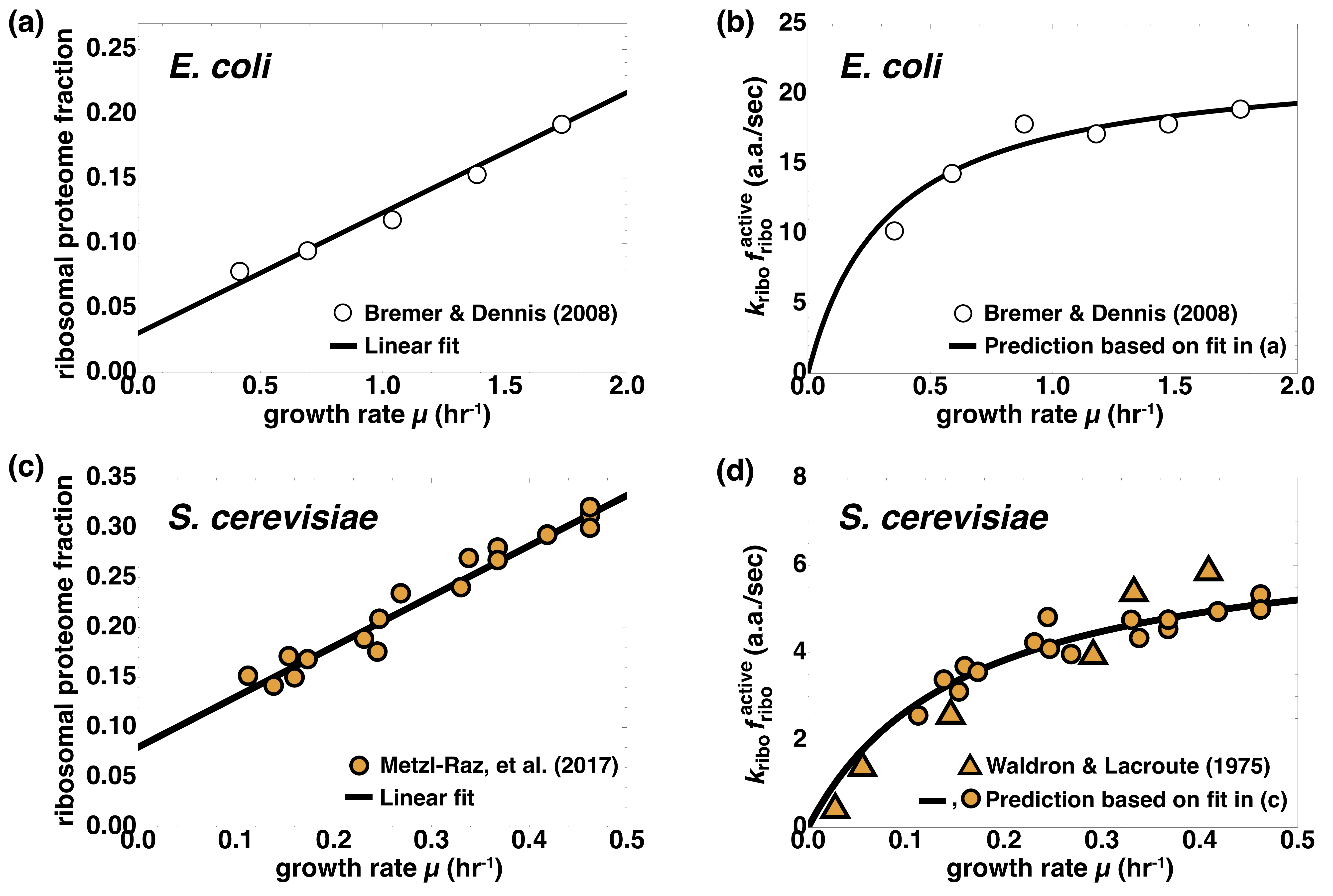}
\vspace{-2mm}
\end{centering}
\caption{\textbf{Growth-law for ribosomal proteome fractions [Eq.~(\ref{growthlaw1})] and Michaelis-Menten behavior of $k_{\text{ribo}} \, f^{\text{active}}_{\text{ribo}}$, in the bacterium \textit{E. coli} (upper panels) and in the eukaryote \textit{S. cerevisiae} (bottom panels)}.  In panel (a), we plot \textit{E. coli} r-protein data from Bremer \& Dennis~\cite{BremerDennis} (circles) and provide a linear fit (solid line): $\phi^{\text{r-prot}}_{\text{ribo}} = 0.093 \mu + 0.031$.  The Michaelis-Menten behavior of $k_{\text{ribo}} \, f^{\text{active}}_{\text{ribo}}$ can be extracted from the linear fit via Eq.~(\ref{MMtoymodel}) to give: $k_{\text{ribo}} \, f^{\text{active}}_{\text{ribo}} \text{ (a.a./sec)} \simeq 22 \mu /(0.33 + \mu)$, where $\mu$ is given in hr$^{-1}$.  This Michaelis-Menten prediction is denoted by the solid line in panel (b), which is in close agreement with experimental values (circles).  The same analysis can be applied to \textit{S. cerevisiae}.  In panel (c), we plot data (orange circles) and a linear fit for r-protein fractions vs. growth rate as provided by Metzl-Raz, et al.~\cite{Barkai}.  We extract the Michaelis-Menten form $k_{\text{ribo}} \, f^{\text{active}}_{\text{ribo}} \text{ (a.a./sec)} \simeq 6.9 \mu /(0.16 + \mu)$ from the linear fit via Eq.~(\ref{MMtoymodel}), as shown in panel (d) by the solid line.  Furthermore, in panel (d) we infer values of $k_{\text{ribo}} \, f^{\text{active}}_{\text{ribo}}$ for each data point in panel (c) using the growth-law in Eq.~(\ref{growthlaw1}).  These predictions appear to be in agreement with measurements by Waldron \& Lacroute~\cite{WaldronLacroute75} (orange triangles) despite the different \textit{S. cerevisiae} strain.}  
\label{rprotgrowthlaw}
\end{figure*}

\subsection{Growth-law for ribosomal protein}

We first consider the proportionality between $\phi^{\text{r-prot}}_{\text{ribo}}$ and growth rate, where we adopt the common interpretation of $\phi^{\text{r-prot}}_{\text{ribo}}$ as the r-protein proteome fraction in the cell [Eq.~(\ref{growthlaw1})].  The same growth-law was shown to hold in the bacterium \textit{E. coli} (Fig.~\ref{rprotgrowthlaw}a).  Plotting $\phi^{\text{r-prot}}_{\text{ribo}}$ vs. $\mu$ as per convention~\cite{Barkai,KlumppPNAS} yields a proportionality factor $N^{\text{a.a.}}_{\text{ribo}}/(k_{\text{ribo}} \, f^{\text{active}}_{\text{ribo}})$.  In principle, both translation rate $k_{\text{ribo}}$ and ribosomal activity $f^{\text{active}}_{\text{ribo}}$ can vary with growth rate.  For example, in \textit{E. coli} $f^{\text{active}}_{\text{ribo}}$ remains constant at 85\% across growth rates, while $k_{\text{ribo}}$ exhibits a Michaelis-Menten dependence characteristic of enzymes, saturating at $\sim\!22$ a.a./sec in rapid growth~\cite{BremerDennis, KlumppPNAS, KR}.  The product $k_{\text{ribo}} \, f^{\text{active}}_{\text{ribo}}$, which appears in the proportionality constant $N^{\text{a.a.}}_{\text{ribo}}/(k_{\text{ribo}} \, f^{\text{active}}_{\text{ribo}})$, then also exhibits a Michaelis-Menten dependence (Fig.~\ref{rprotgrowthlaw}b, circles).  We conjecture that the product of translation rate and ribosome activity also has a Michaelis-Menten form in yeast: 
\begin{equation}
k_{\text{ribo}} \, f^{\text{active}}_{\text{ribo}} = \frac{k^{\text{max}}_{\text{eff}} \mu }{\mu_{\text{HM}}+\mu},
\label{MMform}
\end{equation}
where $k^{\text{max}}_{\text{eff}}$ is the saturation value, and $\mu_{\text{HM}}$ is the growth rate at half its maximum, $k^{\text{max}}_{\text{eff}}/2$.  The Michaelis-Menten dependence manifests itself in the growth-law plots via a non-zero vertical intercept.  To see this, we insert Eq.~(\ref{MMform}) into the growth-law [Eq.~(\ref{growthlaw1})]:
\begin{align}
    \phi^{\text{r-prot}}_{\text{ribo}} & \simeq \frac{N^{\text{a.a.}}_{\text{ribo}}}{k_{\text{ribo}} \, f^{\text{active}}_{\text{ribo}}} \, \mu \simeq N^{\text{a.a.}}_{\text{ribo}} \left( \frac{\mu_{\text{HM}}+\mu}{k^{\text{max}}_{\text{eff}} \mu } \right) \mu \nonumber \\ & \simeq \underbrace{\frac{N^{\text{a.a.}}_{\text{ribo}}}{k^{\text{max}}_{\text{eff}}}}_{\text{slope}} \, \mu + \underbrace{\frac{N^{\text{a.a.}}_{\text{ribo}} \, \mu_{\text{HM}}}{k^{\text{max}}_{\text{eff}}}}_{\text{intercept}}
    \label{MMtoymodel}
\end{align}
Hence the linear dependence on growth rate is preserved as in the case of a constant-valued $k_{\text{ribo}} \, f^{\text{active}}_{\text{ribo}}$, but a non-zero intercept is introduced.  Indeed, a non-zero intercept has been observed in yeast experiments and was interpreted as an excess ribosomal proteome fraction in preparation for increased translation demands when growth conditions change~\cite{Barkai, Waldron77, Koch71}.  In light of Eq.~(\ref{MMtoymodel}), the origin of this non-zero intercept might be traced to a Michaelis-Menten behavior of the ribosomal activity and translation rate product.  

The constants $k^{\text{max}}_{\text{eff}}$ and $\mu_{\text{HM}}$ of the Michaelis-Menten form in Eq.~(\ref{MMform}) can be extracted from a linear fit of $\phi_{\text{ribo}}^{\text{r-prot}}$ vs. $\mu$, via Eq.~(\ref{MMtoymodel}).  As an example, we apply it to \textit{E. coli} data~\cite{BremerDennis} shown in Fig.~\ref{rprotgrowthlaw}a.  We first obtain a linear fit to the data, $\phi_{\text{ribo}}^{\text{r-prot}} \simeq 0.093 \mu + 0.031$, and deduce a Michaelis-Menten behavior of $k_{\text{ribo}} \, f^{\text{active}}_{\text{ribo}} \simeq 22 \mu / (0.33+\mu)$, where we recall that the number of amino acids in the $\textit{E. coli}$ ribosome is 7536~\cite{Bionum}.  As shown in Fig.~\ref{rprotgrowthlaw}b, the fit is in good agreement with data~\cite{BremerDennis}.  

\begin{figure*}[htb]
\begin{centering}
\includegraphics[width=0.75\textwidth]{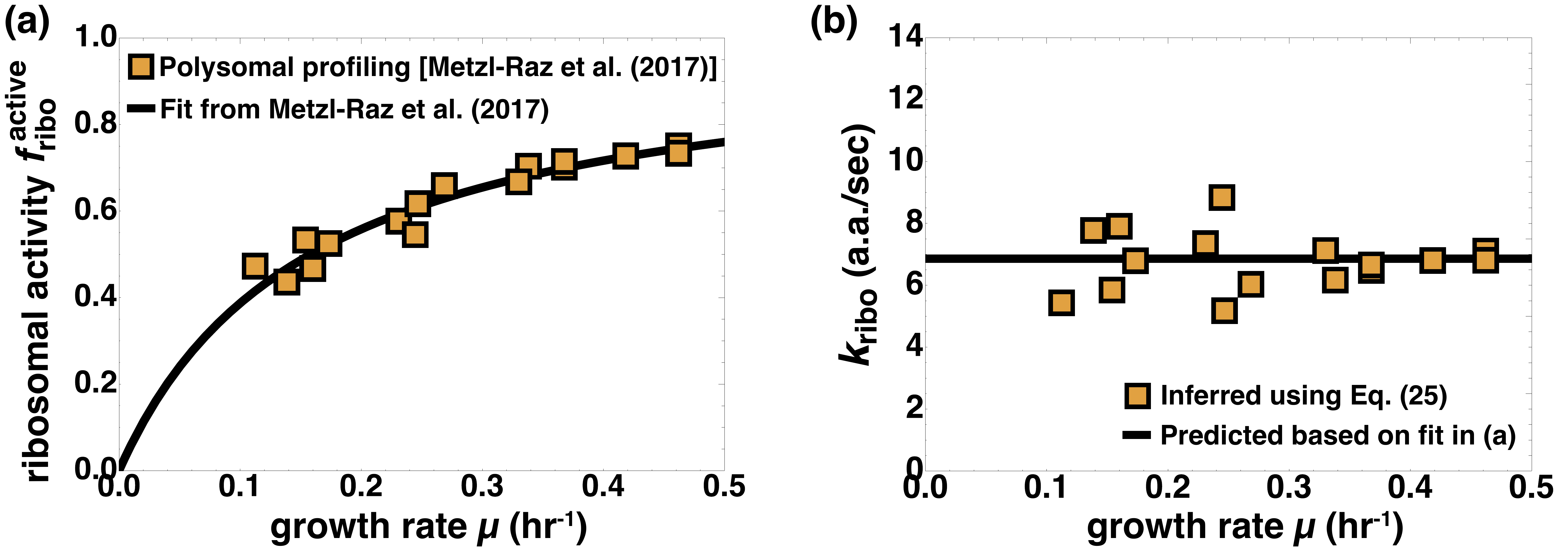}
\vspace{-2mm}
\end{centering}
\caption{\textbf{Inferring the dependence of peptide elongation rate $k_{\text{ribo}}$ on growth rate in yeast.}  In panel (a) we plot polysomal profiling data (orange squares) and the fit $f^{\text{active}}_{\text{ribo}} \simeq \mu / (0.16 + \mu)$ from Fig.~3B of Metzl-Raz, et al.~\cite{Barkai}. The Michaelis-Menten constant of the $f^{\text{active}}_{\text{ribo}}$ fit is the same as that inferred previously for ribosomal efficiency: $k_{\text{ribo}} f^{\text{active}}_{\text{ribo}} \simeq 6.9 \mu / (0.16 + \mu)$.  This implies that the peptide elongation rate $k_{\text{ribo}}$ is approximately constant at 6.9 a.a./sec.  In panel (b) we plot this predicted translation rate (solid line), and compare to values inferred using Eq.~(\ref{growthlaw1}) and r-protein proteome data (Fig.~\ref{rprotgrowthlaw}) from Metzl-Raz, et al.~\cite{Barkai}.}  
\label{inferkribo}
\end{figure*}

We follow the same procedure for \textit{S. cerevisiae} using the data and linear fit reported in Fig.~2A of Ref.~\cite{Barkai}: $\phi_{\text{ribo}}^{\text{r-prot}} \simeq 0.35\mu / \ln(2) + 0.08$ (Fig.~\ref{rprotgrowthlaw}c).  We extracted the following Michaelis-Menten behavior of the effective translation rate in a.a./sec (Fig.~\ref{rprotgrowthlaw}d): $k_{\text{ribo}} \, f^{\text{active}}_{\text{ribo}} \simeq 6.9 \mu /(0.16 + \mu)$, where we used $N^{\text{a.a.}}_{\text{ribo}}=12467$~\cite{SGD} (obtained from a compiled list of ribosomal proteins, see supplemental Excel file).  It implies that the saturation value of $k_{\text{ribo}} \, f^{\text{active}}_{\text{ribo}}$ is $\sim \! 7$ a.a./sec, which is in good agreement with data reported in the literature.  Specifically, cytoplasmic ribosomes in yeast were reported to have average translation rates of $k_{\text{ribo}} \sim 2.8$ to 10.0 a.a./sec~\cite{BoehlkeFriesen,Bonven79,WaldronLacroute75, Piques}. A higher rate of 10.5 a.a./sec was also reported~\cite{Waldron77} under the assumption that translation rate is independent of growth rate, while ribosomal activity varies from 50\%--84\%.  Meanwhile, Bonven \& Gull{\o}v~\cite{Bonven79} reported an active ribosome fraction $f^{\text{active}}_{\text{ribo}}$ of 36\% to 59\%.  An independent study by Metzl-Raz et al.~\cite{Barkai} estimated the active fraction of ribosomes using polysomal profiling, finding it to range from $\sim\!40$\% to 75\% (Fig.~3 of Ref.~\cite{Barkai}).  The maximum values reported for $k_{\text{ribo}} \sim \! 10.0$ a.a./sec and $f^{\text{active}}_{\text{ribo}} \sim \! 75\%$ thus yields a product $k_{\text{ribo}} \, f^{\text{active}}_{\text{ribo}} \sim \! 7.5$ a.a./sec, which is in close agreement with the saturation value of $\sim \! 7$ a.a./sec obtained above.  Furthermore, values of the ``ribosomal efficiency'' $k_{\text{ribo}} \, f^{\text{active}}_{\text{ribo}}$ measured by Waldron 
\& Lacroute~\cite{WaldronLacroute75}, albeit for a different 
\textit{S. cerevisiae} strain, appear to  follow the same Michaelis-Menten trend (triangular markers in Fig.~\ref{rprotgrowthlaw}d).

\subsection{Inferring the dependence of translation rate on growth rate in yeast}

The product $k_{\text{ribo}} \, f^{\text{active}}_{\text{ribo}}$ of ribosomal activity and translation rate appears to exhibit a Michaelis-Menten dependence on growth rate.  However, their separate behaviors, i.e. $k_{\text{ribo}}$ vs. $\mu$ and $f^{\text{active}}_{\text{ribo}}$ vs. $\mu$, are less clear.  Does ribosomal activity in yeast remain constant while translation rate depends on growth rate in a Michaelis-Menten fashion, as in \textit{E. coli}?  Indeed, there are conflicting reports in the literature on \textit{S. cerevisiae}: Waldron et al.~\cite{Waldron77} reported constant translation rates but varying ribosomal activity, while Bonven \& Gulløv~\cite{Bonven79} found that both translation rates and ribosomal activity vary with growth rates.  Meanwhile Boehlke \& Friesen~\cite{BoehlkeFriesen} also found $k_{\text{ribo}}$ to vary with growth rate, but assumed a ribosomal activity of 90\%.  More recently, Metzl-Raz et al.~\cite{Barkai} used polysomal profiling to estimate the active fraction $f^{\text{active}}_{\text{ribo}}$ of ribosomes (Fig.~\ref{inferkribo}a), where monosomes were assumed to be inactive.  They found the Michaelis-Menten behavior $f^{\text{active}}_{\text{ribo}} \simeq \mu/(0.16+\mu)$.  Note that the Michaelis-Menten constant, 0.16, is the same as that of the product $k_{\text{ribo}} \, f^{\text{active}}_{\text{ribo}} \simeq 6.9 \mu/(0.16+\mu)$ extracted in Fig.~\ref{rprotgrowthlaw}.  It follows that $k_{\text{ribo}} \approx 6.9$ a.a./sec (Fig.~\ref{inferkribo}b).  Thus translation rate appears to remain approximately constant across growth rates, as reported by Waldron, et al.~\cite{Waldron77}.

\subsection{Growth-law for RNA polymerases I}

\begin{figure*}[htb]
\begin{centering}
\includegraphics[width=0.8\textwidth]{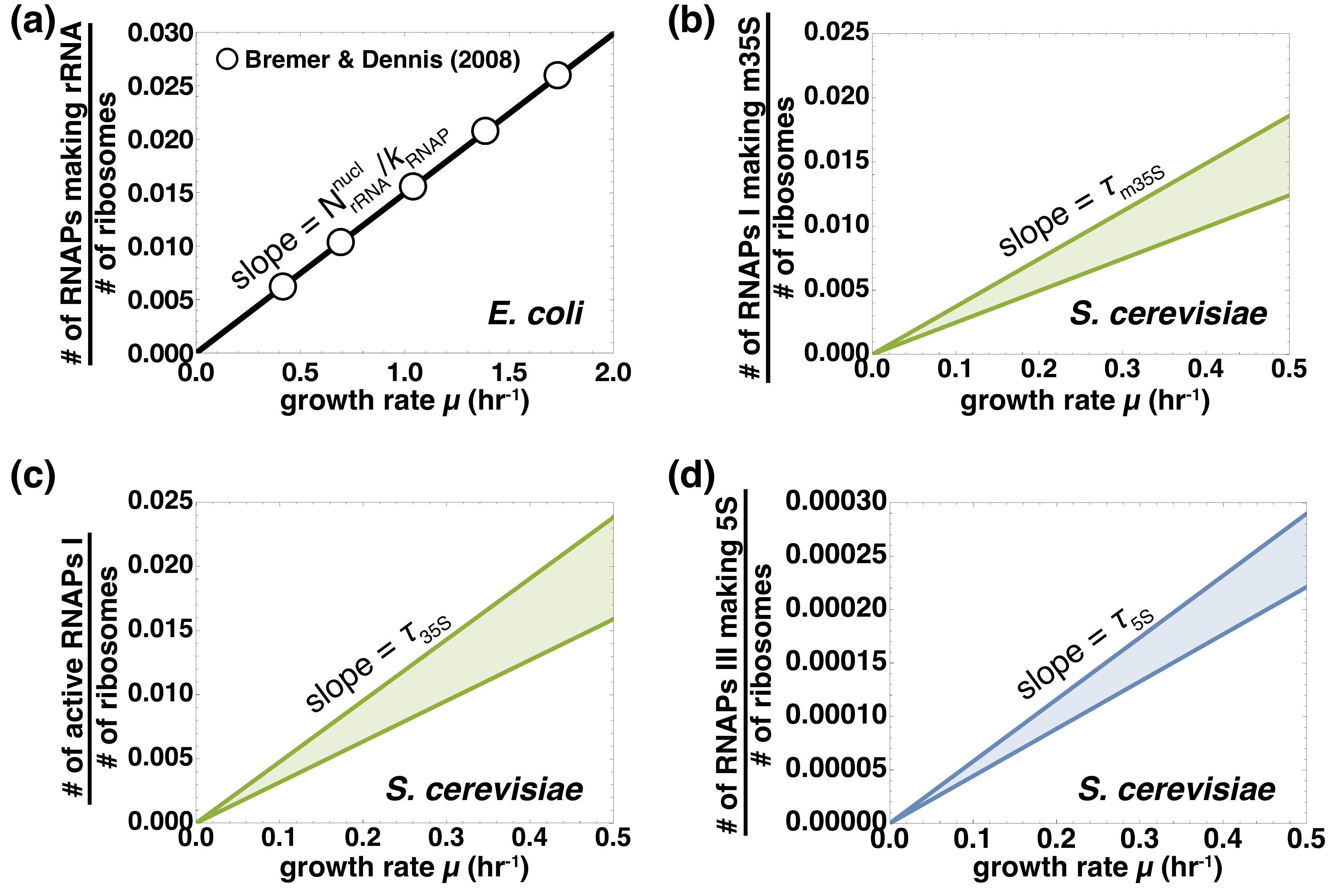}
\vspace{-2mm}
\end{centering}
\caption{\textbf{Predictive growth-laws for the number of RNA polymerases making rRNA relative to the number of ribosomes, assuming constant transcription rates.}  In panel (a) we plot the ratio of the number of RNA polymerases making rRNA to the number of ribosomes in the bacterium \textit{E. coli}.  Transcription rates were observed to stay constant at $\sim \! 85$ nt/sec across growth conditions~\cite{BremerDennis}.  The slope of the solid line, which is a bacterial growth-law equivalent to Eqs.~(\ref{growthlaw2interpretation})--(\ref{growthlaw3interpretation}), is the time required for one RNA polymerase to make a full set of rRNA.  The growth-law (solid line) is in excellent agreement with data (circles).  In panels (b)--(d), we plot these growth-laws for yeast [Eqs.~(\ref{growthlaw2interpretation})--(\ref{growthlaw3interpretation})].  The slope in panel (b) is given by the time for an RNAP I to transcribe a full set of mature 35S-derived rRNAs, i.e. the 18S, 25S, and 5.8S rRNAs [Eq.~(\ref{growthlaw2interpretation})].  In panel (c) the slope is the time required for an RNAP I to transcribe a 35S pre-rRNA, including spacer nucleotides, and thus we obtain the ratio between the number of active RNAPs I and ribosomes [Eq.~(\ref{growthlaw2interpretation2})].  Plotting the last growth-law [Eq.~(\ref{growthlaw3interpretation})] in panel (d), the slope is the time it takes an RNAP III to transcribe a 5S rRNA.  Transcription rates were assumed to remain constant with respect to growth rate, in analogy to \textit{E. coli}.  Reported values at 30$^{\circ}$C range between $k_{\text{RPI}} \sim 40 - 60$ nt/sec~\cite{Kos, FrenchRNAPIrates} and $k_{\text{RPIII}} \sim 58-76$ nt/sec~\cite{FrenchRNAPIII}; the confidence bounds correspond to the maximum and minimum values.}
\label{RNAPgrowthlaws}
\end{figure*}

A similar analysis can be done for the growth-law involving RNAP I-protein [Eqs.~(\ref{growthlaw2interpretation}),  (\ref{growthlaw2interpretation2})].  However, current data for yeast are insufficient to determine how RNAP I transcription rate and activity depend on growth rate.  In \textit{E. coli}, the behaviors of transcription rate and RNAP activity are reversed compared to their translation counterparts: It is RNAP activity which varies with growth rate and saturates at 31\%, while the rRNA transcription rate stays constant at 85 nt/sec across growth conditions (Fig.~\ref{RNAPgrowthlaws}a)~\cite{BremerDennis}.  In analogy to \textit{E. coli}, we plot the growth-law of Eq.~(\ref{growthlaw2interpretation}) for the case of a constant transcription rate, thereby embedding all variability in RNAP I activity.  (Should $k_{\text{RPI}}$ vary in a Michaelis-Menten fashion with growth rate, a non-zero intercept will appear as in the case of r-protein.)  Experiments by French, et al.~\cite{FrenchRNAPIrates} on \textit{S. cerevisiae} grown in a YPD medium at 30$^{\circ}$C -- the same temperature as in Metzl-Raz, et al.~\cite{Barkai} experiments -- indicate RNAP I transcription rates of $k_{\text{RPI}} \sim 54$ to 60 nt/sec for a doubling time of 100 min.  Ko{\v s} \& Tollervey report slightly lower transcription rates of 40 nt/sec at 30$^{\circ}$C~\cite{Kos}.  However, Ko{\v s} \& Tollervey used synthetic growth media which have significantly longer doubling times of $\sim \! 140$ min as compared to $\sim \! 90$ min for YPD growth media~\cite{Sherman}.  This may indicate that there is indeed some dependence of transcription rates on growth rate, but given the absence of more extensive data, we assume the simplistic picture of a constant transcription rate and present the full range of reported rates via the confidence bounds in Fig.~\ref{RNAPgrowthlaws}b.  Note that we have used $N^{\text{nucl}}_{\text{m35S}} = 5354$~\cite{Woolford,Melnikov} (Appendix~\ref{E}), which appears in $\tau_{\text{m35S}}$ on the left-hand side of Eq.~(\ref{growthlaw2interpretation}).  We also provide an alternate form of the growth-law  [Eq.~(\ref{growthlaw2interpretation2})] for the number of active RNAPs I (Fig.~\ref{RNAPgrowthlaws}c), where the number of nucleotides in each 35S pre-rRNA (including spacer nucleotides) is $N^{\text{nucl}}_{\text{35S}} = 6858$~\cite{SGD} (Appendix~\ref{E}).

\subsection{Growth-law for RNA polymerases III}

The remaining growth-law [Eq.~(\ref{growthlaw3interpretation})] for the ratio between the number of RNAPs III making 5S rRNA and the number of ribosomes, vs. growth rate, is plotted in Fig.~\ref{RNAPgrowthlaws}d.  The proportionality factor is $\tau_{\text{5S}} = N^{\text{nucl}}_{\text{5S}}/k_{\text{RPIII}}$, where $N^{\text{nucl}}_{\text{5S}} = 121$~\cite{Woolford, SGD, Melnikov}.  As before, we assume a constant-valued transcription rate, spanning the range $k_{\text{RPIII}} \! \sim \! 58-76$ nt/sec reported by French, et al.~\cite{FrenchRNAPIII} for yeast grown at 30$^{\circ}$C using YPD medium.  

\subsection{How many RNAPs I per RNAP III are required for rRNA production?}

Upon dividing the second growth-law by the third, we obtain the the number of RNAPs I making 18S/25S/5.8S per RNAP III making 5S rRNA.  Or, if using the alternate version of the growth-law in Eq.~(\ref{growthlaw2interpretation2}), we obtain the number of active RNAPs I per RNAP III making 5S.  To estimate their numerical values, we use the nominal values of transcription rates $k_{\text{RPI}} \approx 60$ nt/sec and $k_{\text{RPIII}} \approx 61$ nt/sec reported by French, et al.~\cite{FrenchRNAPIrates, FrenchRNAPIII}.  The characteristic timescales are then $\tau_{\text{m35S}} \approx 89$ sec and $\tau_{\text{5S}} \approx 2.0$ sec.  An exponentially growing yeast cell in YPD medium at 30$^{\circ}$C is therefore predicted to have approximately $\tau_{\text{m35S}}/\tau_{\text{5S}} \approx 45$ RNAPs I making 18S/25S/5.8S per RNAP III synthesizing 5S rRNA.   Equivalently, if including 35S spacer nucleotides, it takes an RNAP I about $\tau_{\text{35S}} \approx 114$ sec to transcribe a full 35S pre-rRNA (6858 nts).  Hence, we find there are $\tau_{\text{35S}}/\tau_{\text{5S}} \approx 57$ active RNAPs I per RNAP III making 5S rRNA.  These numbers can be compared to measurements by French, et al.~\cite{FrenchRNAPIrates, FrenchRNAPIII} in wild-type yeast cells:  The total number of engaged RNAPs I per cell ranged from $\sim \! 3980$ to 4850, with an average of $\sim \! 72$ engaged RNAPs III per cell.  For every RNAP III engaged in 5S synthesis, there are then $\sim \! 55$ to 67 engaged RNAPs I.  This is in close agreement with our theoretical estimate of $\sim \! 57$ active RNAPs I per RNAP III making 5S. 

\subsection{How many ribosomes are in the cell?}

The number of ribosomes per cell can also be inferred from the RNAP growth-laws.  For example, consider Eq.~(\ref{growthlaw2interpretation2}) combined with the numbers given above, i.e. $\sim \! 3980$ to 4850 engaged RNAPs I per cell, and a 35S transcription timescale of $\tau_{\text{35S}} \approx 114$ sec.  Assuming doubling times of $\approx \! 100$ min for yeast in YPD media at $30^{\circ}$C~\cite{Warner, WaldronLacroute75}, we find the number of ribosomes per cell to lie in the range 301,400 to 367,300.  While this range may seem high compared to the 200,000 estimate provided by one source~\cite{Warner}, it agrees well with measurements of $\sim \! 348,000$ ribosomes/cell by Waldron \& Lacroute~\cite{WaldronLacroute75}.  Note that such estimates decrease for faster growth rates (assuming the same number of RNAPs), e.g. for doubling times of 90 min instead of 100 min we find a range of $\sim \! 271,300$ to 330,600 ribosomes/cell.

A similar estimate can be obtained via the RNAP III growth-law [Eq.~(\ref{growthlaw3interpretation})].  Recall the 5S transcription timescale $\tau_{\text{5S}} \approx 2.0$ sec and the estimate of $\sim \! 72$ engaged RNAPs III per cell.  For a doubling time of $\sim \! 100$ minutes, we then obtain an estimate of $\sim \! 314,200$ ribosomes/cell, which is consistent with the range predicted by the RNAP I growth-law.  Conversely, one could obtain estimates for the number of RNAPs I and III making rRNA in the cell, based on measurements of the number of ribosomes per cell.

\subsection{Future outlook: Using the invariants to deduce activities of RNA polymerases I and III from their proteome fractions}

\begin{figure*}[htb]
\begin{centering}
\includegraphics[width=\textwidth]{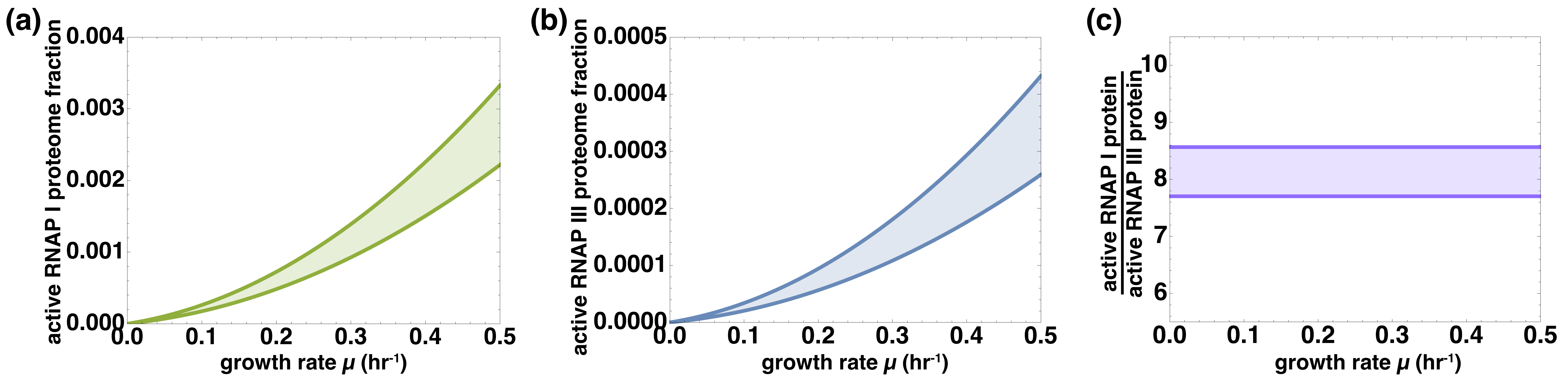}
\vspace{-5mm}
\end{centering}
\caption{\textbf{Predicted proteome fractions of active RNAPs I and RNAPs III in yeast.}  Panel (a) displays the predicted proteome fraction of active RNAPs I, $f^{\text{active}}_{\text{RPI}} \phi^{\text{RPI}}_{\text{ribo}}$, vs. growth rate according to Eq.~(\ref{growthlaw2}) and the ribosomal proteome fraction fit from Ref.~\cite{Barkai}: $\phi^{\text{r-prot}}_{\text{ribo}} 
\simeq (0.35/\ln 2) \mu + 0.08$.  We assume here that the RNAP III transcription rate is constant; the confidence bounds correspond to its reported range of $k_{\text{RPI}} \sim 40-60$ nt/sec.  Similarly, in panel (b) we plot the predicted proteome fraction of active RNAPs III, $f^{\text{active}}_{\text{RPIII}} \phi^{\text{RPIII}}_{\text{ribo}}$, vs. growth rate using Eq.~(\ref{growthlaw3}) and the same fit for the ribosomal proteome fraction.  We assume a constant RNAP III transcription rate in the range $k_{\text{RPIII}} \sim 58-76$ nt/sec, and $\phi^{\text{5S}}_{\text{RPIII}} \approx 0.11-0.14$, where lower values in the latter correspond to lower growth rates since there are then more tRNAs per ribosome~\cite{WaldronLacroute75}.  RNAP I activity ($f^{\text{active}}_{\text{RPI}}$) and RNAP III activity ($f^{\text{active}}_{\text{RPIII}}$) can be deduced once the RNAP I and RNAP III proteome fractions are known.  Under the noted assumptions, the active RNAP I and active RNAP III proteome fractions feature a quadratic dependence on the growth rate $\mu$, but their ratio is constant at $\approx 8$ as shown in panel (c).}  
\label{activeRPfractions}
\end{figure*}

Large-scale proteomics studies allow for the estimation of various proteome fractions in the cell.  While there is still large variability in current state-of-the-art proteomics studies~\cite{UnificationProt}, in principle such data can be compared to our predicted values for proteome fractions of ribosomes, and of RNAPs I and III making rRNA.  We outline a method below to extract RNAP I and III activities which, to our knowledge, have not yet been reported.  This method can be used in the near future as more accurate proteomics measurements become available. 

The numerical values of the invariants are given by the true ribosome composition, as per the left-hand sides of Eqs.~(\ref{inv1}) and (\ref{inv2}).  Their values are determined by the protein and rRNA masses of the \textit{S. cerevisiae} ribosome: Each ribosome is composed of 1.40 MDa protein (supplementary Excel sheet) and 1.79 MDa rRNA ~\cite{SGD}.  The rRNA mass was obtained from nucleotide sequences of 18S (587.0 kDa), 25S (1109.7 kDa), 5.8S (51.4 kDa), and 5S (39.4 kDa) mature rRNAs~\cite{SGD}.  This yields a total ribosome mass of 3.2 MDa, such that $x_{\text{prot}} \approx 0.439$ and $x_{\text{m35S}} \approx 0.548$.  The numerical value of the first invariant [Eq.~(\ref{inv1})] is then $x_{\text{prot}}/x_{\text{m35S}} \approx 0.80$, while that of the second [Eq.~(\ref{inv2})] is $x_{\text{m35S}}/(1-x_{\text{prot}}-x_{\text{m35S}}) \approx 44.388$.

Upon examining the right-hand side of the first invariant in Eq.~(\ref{inv1}), all but the rightmost ratio is known.  The values described earlier for RNAP I transcription rate $k_{\text{RPI}}$ and peptide elongation rate $k_{\text{ribo}}$ can be used in Eq.~(\ref{inv1}).  The number of amino acids in the RNAP I is $N^{\text{a.a.}}_{\text{RPI}}=5236$, while the ribosome has $N^{\text{a.a.}}_{\text{ribo}}=12467$ amino acids (Appendix~\ref{E}, supplementary Excel file)~\cite{SGD}.  We estimate the average amino acid and nucleotide masses as $m_{\text{a.a.}} \approx 112$ Da and $m_{\text{nucl}} \approx 326$ Da, based on the composition of the ribosome and three RNA polymerases.  While these values require fine tuning to reflect all amino acids and nucleotides in the cell, they are already in good agreement with the average \textit{E. coli} amino acid mass (109 Da) and nucleotide mass (324.3 Da)~\cite{Bionum}.  

Returning to the right-hand side of Eq.~(\ref{inv1}), we can also deduce the fraction $\phi^{\text{m35S}}_{\text{RPI}}$ of active RNAPs I which synthesize mature rRNAs as opposed to 35S spacer nucleotides (Fig.~\ref{rRNAprocessing}).  Accounting for spacer nucleotides in pre-rRNA was shown to be critical in \textit{E. coli}~\cite{KR}.  In yeast, the 35S pre-rRNA contains a total of 1504 spacer nucleotides from: ITS1 (361 nts), ITS2 (232 nts), 5'ETS (700 nts), and 3'ETS (211 nts)~\cite{SGD}, where ITS and ETS denote an internally transcribed spacer and an externally transcribed spacer, respectively.  Including these spacer nucleotides yields a total of 6858 nucleotides in each 35S pre-rRNA.  We therefore estimate the fraction of active RNAPs I dedicated to the transcription of mature rRNAs as $\phi^{\text{m35S}}_{\text{RPI}} \simeq 5354/6858 \simeq 78\%$.

Remaining on the right-hand of Eq.~(\ref{inv1}) are the proteome fractions $\phi^{\text{r-prot}}_{\text{ribo}}$ and $\phi^{\text{RPI}}_{\text{ribo}}$, ribosomal activity $f^{\text{active}}_{\text{ribo}}$, and RNAP I activity $f^{\text{active}}_{\text{RPI}}$.  As discussed earlier, the ribosomal activity can be estimated using the Michaelis-Menten dependence on growth rate shown in Fig.~\ref{inferkribo}.  It follows that RNAP I activity can be deduced once the ribosomal and RNAP I proteome fractions are known.  

RNAP III activity can be determined from the second invariant [Eq.~(\ref{inv2})] in a similar fashion.  On the right-hand side, we have the number of amino acids $N^{\text{a.a.}}_{\text{RPI}} = 5236$ and $N^{\text{a.a.}}_{\text{RPIII}} = 6151$ in RNAP I and III, respectively~\cite{SGD}.  Ranges of the RNAP I and III transcription rates $k_{\text{RPI}}$ and $k_{\text{RPIII}}$, mentioned in a previous section, are provided by French et al.~\cite{FrenchRNAPIrates,FrenchRNAPIII}.  The quantity $\phi^{\text{m35S}}_{\text{RPI}} \approx 78\%$ is also known.  Thus, aside from RNAP I and III activities and proteome fractions, remaining is the fraction $\phi^{\text{5S}}_{\text{RPIII}}$ of active RNAPs III which synthesize 5S rRNA.  This quantity can be estimated using measurements for the number of tRNAs per ribosome in the cell~\cite{WaldronLacroute75}.  RNAPs III synthesizes the 5S rRNA, nuclear tRNAs, and a few other small nuclear RNAs whose contribution we henceforth neglect~\cite{SGD}.  Waldron \& Lacroute found that there are about 9.5 to 12.2 tRNAs per ribosome in the \textit{S. cerevisiae} cell, depending on growth rate.  The average length of a tRNA is 80 nucleotides (supplementary Excel file)~\cite{SGD}.  Thus, for each 5S rRNA (121 nt long), an active RNAP III synthesizes 760 to 976 tRNA nucleotides.  We therefore estimate that $\phi^{\text{5S}}_{\text{RPIII}} \approx 11\%$ to 14\% of active RNAPs III synthesize 5S rRNA.  

Assuming that RNAP I activity was deduced from the first invariant  [Eq.~(\ref{inv1})] as described above, only RNAP I and RNAP III proteome fractions are needed to determine RNAP III activity.  Alternatively, if RNAP I activity is not known, the RNAP III activity can still be extracted using ribosomal activity and the ribosomal proteome fraction: RNAP I activity is altogether eliminated from the second invariant [Eq.~(\ref{inv2})] upon multiplication with the first [Eq.~(\ref{inv1})]. 

Lastly, in Fig.~\ref{activeRPfractions} we illustrate the predicted proteome fraction of active RNAPs I ($f^{\text{active}}_{\text{RPI}} \phi^{\text{RPI}}_{\text{ribo}}$) and of active RNAPs III ($f^{\text{active}}_{\text{RPIII}} \phi^{\text{RPIII}}_{\text{ribo}}$) using the growth-laws [Eqs.~(\ref{growthlaw2}), (\ref{growthlaw3})].  There we assume a ribosomal proteome fraction as given by the fit in Ref.~\cite{Barkai}: $\phi^{\text{r-prot}}_{\text{ribo}} 
\simeq (0.35/\ln 2) \mu + 0.08$.  We also assume constant RNAP I and RNAP III transcription rates which lie in the ranges $k_{\text{RPI}} \sim 40-60$ nt/sec and $k_{\text{RPIII}} \sim 58-76$ nt/sec.  Once the RNAP I and RNAP III proteome fractions are known, RNAP I and RNAP III activities can be readily extracted.

\section{Concluding remarks}\label{conc}

In this work, we presented a kinetic analysis of ribosome biogenesis for lower Eukarya in balanced exponential growth. Three growth-laws and two invariants, akin to those found for Bacteria earlier this year~\cite{KR}, were derived. The first growth-law establishes a proportionality between the cellular growth rate and the proteome mass fraction of r-protein. This proportionality has already been observed in yeast~\cite{KiefWarner, Barkai}, allowing for the inference of a Michaelis-Menten behavior of the ``ribosomal efficiency,'' i.e. the product of ribosomal activity and peptide elongation rate.  The inferred dependence on growth rate was then shown to be in good agreement with an independent set of measurements, despite the use of a different yeast strain~\cite{WaldronLacroute75}. The second and third growth-laws, which yield the number of RNAPs I and III making rRNA per ribosome, also appear to be in good agreement with measurements thus far~\cite{FrenchRNAPIrates, FrenchRNAPIII}. These results suggest that Bacteria and lower Eukarya obey similar growth-laws despite differences in cellular organization and complexity. 

Because a comprehensive eukaryotic dataset is lacking in the literature, several predictions from our analysis still require verification. Noteworthy in that regard are the timescales \{$\tau_{\text{r-prot}},\tau_{\text{m35S}},\tau_{\text{5S}}$\} appearing in the growth-laws [Eqs.~(\ref{growthlaw1interpretation}), (\ref{growthlaw2interpretation}), (\ref{growthlaw3interpretation})], which couple translation and transcription rates to cell physiology via the ribosome composition.  Their values are consistent with available data for a limited set of growth conditions. However, concurrent measurements of translation and transcription rates \textit{in vivo} are necessary to fully corroborate the growth-laws in light of the microscopic interpretation of the timescales involved.
The predicted invariant quantities of eukaryotic growth, given by Eqs.~(\ref{inv1}) and (\ref{inv2}), also await experimental verification.  Together with the growth-laws, these invariants could eventually be used as a proxy for direct measurements of various kinetic and physiological parameters in eukaryotic cells.  For example, they can be used to infer values of RNAP activity, which have not yet been measured.  A method to deduce such parameters was outlined in the previous section, where we applied the invariants to \textit{S. cerevisiae}.

Since the kinetic analysis presented here relies only on the assumptions of balanced exponential growth and growth rate maximization, the relations we have derived are likely to hold for species other than budding yeast. This would indicate that the ribosome composition in such organisms is tuned to maximize cellular growth rates, as was already verified for \textit{E. coli} \cite{KR} but remains to be confirmed for \textit{S. cerevisiae} and other microorganisms.  Furthermore, because there is some variation in cytoplasmic ribosome composition amongst Eukarya, the relations derived herein might help advance our understanding of ribosome heterogeneity and its consequences~\cite{Moll}.  Specifically, the invariants imply that ribosome composition [left-hand sides of Eqs.~(\ref{inv1}), (\ref{inv2})] is directly coupled to cell physiology [right-hand sides of Eqs.~(\ref{inv1}), (\ref{inv2})]. Yet, how the latter changes to accommodate for different ribosome compositions, i.e. via changes in proteome fractions, translation, or transcription kinetics, remains an open question. 
Finally, it would be interesting to see whether similar growth-laws and invariants hold for higher, more evolved Eukarya, as these share many features of ribosome biogenesis with lower Eukarya and the eukaryotic core proteome appears to be quite stable across species~\cite{EukaryaProteomeStable}. Particularly interesting in this regard are cancer cells which exhibit rapid cell proliferation like Bacteria and yeast~\cite{Dai, Thomson}. Hyperactivated ribosome production is a known signature of rapidly proliferating cancer cells~\cite{Dai,Pelletier}. During tumorigenesis, excessive rRNA transcription leads to enlarged nucleoli, which are the primary sites of ribosome biogenesis in the eukaryotic cell~\cite{White}. Consequently,  nucleolar size in cancer tissues is sometimes used as an indicator of the severity of the disease~\cite{CancerNucleoli}. A more quantitative understanding of ribosome biogenesis would advance cancer research and our understanding of tumorigenesis.  To this end, the analysis presented herein may aid in the search for cancer cell growth-laws. 

\subsection*{Acknowledgments}
The authors thank Eyal Metzl-Raz and Naama Barkai for their help with polysomal profiling and proteomics data.  S.K. thanks Elizabeth R. Chen for helpful advice on the figures.  S.K. was supported by the Zuckerman STEM postdoctoral fellowship.  S.R. acknowledges support from the Azrieli Foundation, from the Raymond and Beverly Sackler Center for Computational Molecular and Materials Science at Tel Aviv University, and from the Israel Science Foundation (Grant No. 394/19).   
\appendix 
\section{Derivation of parametric equations for intersection of bounds in Fig.~\ref{surfaces} and optimal values of $\{x_{\text{prot}}, x_{\text{m35S}}, \mu\}$}\label{A}

From the bounds given in Eqs.~(\ref{bound1xy}), (\ref{bound2xy}), (\ref{bound3xy}) of the main text, we see that the first behaves as $\mu \simeq a/x_{\text{prot}}$, the second as $\mu \simeq b/\sqrt{x_{\text{m35S}}}$, and the third as $\mu \simeq c/\sqrt{1-x_{\text{prot}}-x_{\text{m35S}}}$, where 
\begin{align}
    & a = \frac{m_{\text{a.a.}} \, k_{\text{ribo}} \, \phi_{\text{ribo}}^{\text{r-prot}} \, f^{\text{active}}_{\text{ribo}}}{M_{\text{ribo}}} \, , \\
    & b = \sqrt{\frac{m_{\text{nucl}}}{M_{\text{ribo}} \,  N^{\text{a.a.}}_{\text{RPI}}}  \cdot k_{\text{RPI}} \, \phi^{\text{m35S}}_{\text{RPI}} \, f^{\text{active}}_{\text{RPI}} \cdot k_{\text{ribo}} \, \phi^{\text{RPI}}_{\text{ribo}} \, f^{\text{active}}_{\text{ribo}}} \, , \\
    & c = \sqrt{\frac{m_{\text{nucl}}}{M_{\text{ribo}} \, N^{\text{a.a.}}_{\text{RPIII}}} \cdot k_{\text{RPIII}} \, \phi^{\text{5S}}_{\text{RPIII}} \, f^{\text{active}}_{\text{RPIII}} \cdot  k_{\text{ribo}} \, \phi^{\text{RPIII}}_{\text{ribo}} \, f^{\text{active}}_{\text{ribo}}} \, .
\end{align}
To find the parametric equations for their lines of intersections, we equate every possible pair of bounds.  One line is found by equating the bounds of r-protein and mature 35S-derived rRNA production, from which one obtains $x_{\text{m35S}} = (b x_{\text{prot}}/a)^{2}$.  Therefore the first intersection line can be written parametrically in terms of $x_{\text{prot}}$ as
\begin{equation}
    \text{Line 1: } \{ x_{\text{prot}},x_{\text{m35S}},\mu\}=\left\{x_{\text{prot}}, \, \left(\frac{b}{a} x_{\text{prot}} \right)^{2}, \, \frac{a}{x_{\text{prot}}} \right\} 
    \label{line1}
\end{equation}
It can also be expressed solely in terms of $x_{\text{m35S}}$ if desired.  Similarly, equating the bounds from mature 35S-derived and 5S rRNA production yields $x_{\text{m35S}} = (1-x_{\text{prot}})/(1+c^{2}/b^{2})$ and the corresponding parametric equation
\begin{equation}
    \text{Line 2: } \{ x_{\text{prot}},x_{\text{m35S}},\mu\}=\left\{x_{\text{prot}}, \, \frac{1-x_{\text{prot}}}{1+c^{2}/b^{2}}, \, b \, \sqrt{\frac{1+c^{2}/b^{2}}{1-x_{\text{prot}}}} \right\} \, .
    \label{line2}
\end{equation}
Finally, equating the bounds of r-protein and 5S rRNA production gives $x_{\text{m35S}}=1-x_{\text{prot}}-(c x_{\text{prot}}/a)^{2}$ and the parametric equation
\begin{equation}
    \text{Line 3: } \{ x_{\text{prot}},x_{\text{m35S}},\mu\}=\left\{x_{\text{prot}}, \, 1-x_{\text{prot}}-\left(\frac{c}{a} \, x_{\text{prot}} \right)^{2}, \, \frac{a}{x_{\text{prot}}} \right\} \, .
    \label{line3}
\end{equation}
All three lines (and thus all three surfaces) intersect at one point, which can be seen upon equating every pair of equations in Eqs.~(\ref{line1})--(\ref{line3}) and solving a quadratic equation for $x_{\text{prot}}$.  Only one solution of $x_{\text{prot}}$ is positive and thus physically realizable.  Because of the monotonic behavior of the bounds, this point gives the maximal possible growth rate $\mu_{\text{opt}}$ and corresponding optimal ribosome composition, i.e. optimal r-protein and mature 35S-derived rRNA mass fractions, which we denote here as $x^{*}_{\text{prot}}$ and $x^{*}_{\text{m35S}}$.  The point of intersection and optimal values can be then expressed as
\begin{equation}
\begin{cases}
x^{*}_{\text{prot}} = \frac{-1 + \sqrt{1 + 4\left( \frac{b^2}{a^2} + \frac{c^2}{a^2} \right) } }{ 2 \left( \frac{b^2}{a^2} + \frac{c^2}{a^2} \right) } \, \\
x^{*}_{\text{m35S}} = \frac{-1 + \sqrt{1 + 4\left( \frac{b^2}{a^2} + \frac{c^2}{a^2} \right) }}{2 \left( 1 + \frac{c^2}{b^2} \right) } \, \\
\mu_{\text{opt}} = \frac{2 \left( \frac{b^2}{a} + \frac{c^2}{a} \right) }{-1 + \sqrt{1 + 4\left( \frac{b^2}{a^2} + \frac{c^2}{a^2} \right) }}.
\end{cases}
\end{equation}

\section{Parameter values used in Fig.~\ref{surfaces} of the main text}\label{B}
Parameter values were chosen for the purpose of clearly illustrating the three bounds and their intersections: $f^{\text{active}}_{\text{ribo}} = 0.70$; $f^{\text{active}}_{\text{RPI}} = 0.060$;  $f^{\text{active}}_{\text{RPIII}} = 0.89$; $k_{\text{ribo}} = 26450$ a.a./hr; $k_{\text{RPI}} = 180400$ nucl/hr; $k_{\text{RPIII}} = 93450$ nucl/hr;  $\phi^{\text{m35S}}_{\text{RPI}} = 0.736$; $\phi^{\text{5S}}_{\text{RPIII}} = 0.669$; $\phi^{\text{r-prot}}_{\text{ribo}} = 0.10555$;  $\phi^{\text{RPI}}_{\text{ribo}} = 0.005845$; $\phi^{\text{RPIII}}_{\text{ribo}} = 0.000798$.  This gives the coefficients $a=0.066$, $b=0.13$, $c=0.12$ in Eqs.~(\ref{line1}), (\ref{line2}), and (\ref{line3}).  The point of intersection is thus $\{ x^{*}_{\text{prot}}, \, x^{*}_{\text{m35S}}, \, \mu_{\text{opt}} \} = \{0.31, 0.38, 0.21\}$.

\section{Alternate form of invariant in Eq.~(\ref{inv1})}\label{C}

To obtain an alternative to the invariant presented in Eq.~(\ref{inv1}) of the main text, consider dividing the first growth-law [Eq.~(\ref{growthlaw1})] by the third [Eq.~(\ref{growthlaw3})], and multiplying numerator and denominator by $m_{\text{nucl}} \, m_{\text{a.a.}}$ on the left-hand side.  Recognizing that $m_{\text{a.a.}} N^{\text{a.a.}}_{\text{ribo}} = x_{\text{prot}} M_{\text{ribo}}$ and $m_{\text{nucl}} N^{\text{nucl}}_{\text{5S}} = (1-x_{\text{prot}}- x_{\text{m35S}}) M_{\text{ribo}}$, some rearrangement then yields the invariant quantity which is the ratio between the r-protein and 5S rRNA mass fractions:
\begin{equation}
    \frac{x_{\text{prot}}}{1-x_{\text{prot}}-x_{\text{m35S}}} \simeq \frac{m_{\text{a.a.}}}{m_{\text{nucl}}} \cdot \frac{N^{\text{a.a.}}_{\text{RPIII}}}{N^{\text{a.a.}}_{\text{ribo}}} \cdot \frac{k_{\text{ribo}} }{k_{\text{RPIII}}} \cdot \frac{\phi^{\text{r-prot}}_{\text{ribo}} f^{\text{active}}_{\text{ribo}} \phi^{\text{r-prot}}_{\text{ribo}}}{\phi^{\text{5S}}_{\text{RPIII}} \, f^{\text{active}}_{\text{RPIII}} \, \phi^{\text{RPIII}}_{\text{ribo}}}  \, .
    \label{inv1alt}
\end{equation}

\section{Derivation of proteome fraction interpretations of the invariants in Section~\ref{VII}}\label{D}

The first invariant quantity [Eq.~(\ref{inv1}) of the main text], i.e. the ratio between the r-protein mass fraction squared and the 35S-derived rRNA mass fraction, is:
\begin{equation}
    \frac{x_{\text{prot}}}{x_{\text{m35S}}} \simeq \frac{m_{\text{a.a.}}}{m_{\text{nucl}}} \cdot \frac{N^{\text{a.a.}}_{\text{RPI}}}{N^{\text{a.a.}}_{\text{ribo}}} \cdot \frac{k_{\text{ribo}} }{k_{\text{RPI}}} \cdot \frac{\phi^{\text{r-prot}}_{\text{ribo}} f^{\text{active}}_{\text{ribo}} \phi^{\text{r-prot}}_{\text{ribo}}}{\phi^{\text{m35S}}_{\text{RPI}} \, f^{\text{active}}_{\text{RPI}} \, \phi^{\text{RPI}}_{\text{ribo}}} \, .
    \label{inv1supp}
\end{equation}
To obtain Eq.~(\ref{inv1int}) of the main text, first consider the following interpretation using proteome fractions:
\begin{equation} 
    \frac{x_{\text{prot}}}{x_{\text{m35S}}}\simeq \frac{m_{\text{a.a.}}}{m_{\text{nucl}}} \cdot \frac{N^{\text{a.a.}}_{\text{RPI}}}{N^{\text{a.a.}}_{\text{ribo}}} \cdot \frac{k_{\text{ribo}} }{k_{\text{RPI}}} \cdot \frac{\text{r-protein making r-protein}}{\text{RPI-protein making m35S}} \, .
\end{equation}
The proteome fractions can be converted to numbers of macromolecules since each ribosome has a protein mass of $m_{\text{a.a.}} N^{\text{a.a.}}_{\text{ribo}}$, and each RNAP I has a protein mass $m_{\text{a.a.}} N^{\text{a.a.}}_{\text{RPI}}\,$:
\begin{equation}
\frac{x_{\text{prot}}}{x_{\text{m35S}}}\simeq \frac{m_{\text{a.a.}}}{m_{\text{nucl}}} \cdot \frac{N^{\text{a.a.}}_{\text{RPI}}}{N^{\text{a.a.}}_{\text{ribo}}} \cdot \frac{k_{\text{ribo}} }{k_{\text{RPI}}} \cdot \frac{N^{\text{a.a.}}_{\text{ribo}}}{N^{\text{a.a.}}_{\text{RPI}}} \cdot \frac{\text{\# of ribosomes making r-protein}}{\text{\# of RNAPs I making m35S}} \, ,
\end{equation}
which simplifies to Eq.~(\ref{inv1int}) of the main text.

The second invariant quantity [Eq.~(\ref{inv2}) of the main text], the mass ratio between 35S-derived mature rRNAs and 5S rRNA, is:
\begin{equation}
    \frac{x_{\text{m35S}}}{1-x_{\text{prot}}-x_{\text{m35S}}} \simeq \frac{N^{\text{a.a.}}_{\text{RPIII}}}{N^{\text{a.a.}}_{\text{RPI}}} \cdot \frac{k_{\text{RPI}}}{k_{\text{RPIII}}} \cdot \frac{\phi^{\text{m35S}}_{\text{RPI}} \, f^{\text{active}}_{\text{RPI}} \, \phi^{\text{RPI}}_{\text{ribo}}}{\phi^{\text{5S}}_{\text{RPIII}} \, f^{\text{active}}_{\text{RPIII}} \, \phi^{\text{RPIII}}_{\text{ribo}} } \, .
    \label{inv2supp}
\end{equation}
Using the proteome fraction interpretation, we obtain
\begin{equation}
    \frac{x_{\text{m35S}}}{1-x_{\text{prot}}-x_{\text{m35S}}} \simeq \frac{N^{\text{a.a.}}_{\text{RPIII}}}{N^{\text{a.a.}}_{\text{RPI}}} \cdot \frac{k_{\text{RPI}}}{k_{\text{RPIII}}} \cdot \frac{\text{RPI-protein making m35S}}{\text{RPIII-protein making 5S}} \, .
\end{equation}
As before, we convert the protein masses to numbers of macromolecules given that each RNAP I and RNAP III has a protein mass of $m_{\text{a.a.}} N^{\text{a.a.}}_{\text{RPI}}\,$ and $m_{\text{a.a.}} N^{\text{a.a.}}_{\text{RPIII}}\,$, respectively, which yields Eq.~(\ref{inv2int}) of the main text.

\begin{widetext}
\section{Publicly available data for \textit{S. cerevisiae}} \label{E}

\begin{table*}[h!] 
\noindent \begin{centering}
\begin{threeparttable}
\caption{\textbf{Biological parameters having fixed values (independent of growth conditions)}}
\begin{tabular}{|l|c|c|}
\hline 
 & Symbol & Value \tabularnewline
\hline
\hline
Number of amino acids in the ribosome~\cite{SGD}~\tnote{\textit{a}} 
& $N^{\text{a.a.}}_{\text{ribo}}$ & 12467 \tabularnewline
\hline 
Number of nucleotides in the 5S rRNA~\cite{Woolford, SGD, Melnikov} 
& $N^{\text{nucl}}_{\text{5S}}$ & 121 \tabularnewline
\hline
Number of nucleotides in the 25S rRNA~[13,20,21] &  & 3396 \tabularnewline
\hline
Number of nucleotides in the 5.8S rRNA~[13,20,21] &  & 158 \tabularnewline
\hline
Number of nucleotides in the 18S rRNA~[13,20,21] &  & 1800 \tabularnewline
\hline
Number of nucleotides in  mature 35S-derived (25S, 5.8S, 18S) rRNAs &  $N^{\text{nucl}}_{\text{m35S}}$ &  5354 \tabularnewline
\hline
Total number of nucleotides in 35S pre-rRNA, including spacers~[20]~\tnote{\textit{b}} \hspace{1mm} 
& $N^{\text{nucl}}_{\text{35S}}$ & 6858  \tabularnewline
\hline
Ribosome mass~\cite{SGD}~\tnote{\textit{c}}
& $M_{\text{ribo}}$ & 3.2 MDa  \tabularnewline
\hline
Number of amino acids in RNAP I~\cite{SGD}~\tnote{\textit{d}} 
&  $N^{\text{a.a.}}_{\text{RPI}}$ & \multicolumn{1}{c|}{ 5236} \tabularnewline
\hline 
RNAP I mass~\cite{SGD}~\tnote{\textit{d}} & $M_{\text{RPI}}$ & \multicolumn{1}{c|}{ 0.59 MDa } \tabularnewline
\hline 
Number of amino acids in RNAP III~\cite{SGD}~\tnote{\textit{d}}  &  $N^{\text{a.a.}}_{\text{RPIII}}$ & \multicolumn{1}{c|}{ 6151} \tabularnewline
\hline 
RNAP III mass~\cite{SGD}~\tnote{\textit{d}} & $M_{\text{RPIII}}$ & \multicolumn{1}{c|}{ 0.69 MDa } \tabularnewline
\hline
Estimated average mass of an amino acid in the cell~\tnote{\textit{e}} & $m_{\text{a.a.}}$ & $\sim \! 112$ Da \tabularnewline
\hline
Estimated average mass of a nucleotide in the cell~\tnote{\textit{f}} & $m_{\text{nucl}}$ & $\sim \! 326$ Da \tabularnewline
\hline
Average tRNA length~\cite{SGD}~\tnote{\textit{g}} &  & 80 nt  \tabularnewline
\hline
\end{tabular}
\label{tab:fixednum}
  \begin{tablenotes}
    \item[\textit{a}]  See supplementary Excel file for a list of ribosomal protein subunits and their corresponding number of amino acids.
    \item[\textit{b}]  See Fig.~\ref{rRNAprocessing}.
    \item[\textit{c}]  Molecular weights for protein subunits are listed in a supplementary Excel file.  The rRNA molecular weights were calculated from rRNA sequences listed in the SGD database~\cite{SGD}.  Some references report a slightly ribosome mass of 3.3 MDa~\cite{Melnikov}.
    \item[\textit{d}]  See supplementary Excel file for data regarding RNAP I, II, III protein subunits and masses.
    \item[\textit{e}]  Based on molecular weights of ribosomal and RNAP I, II, and III proteins, see supplementary Excel file.  This value is slightly higher than that reported in \textit{E. coli} (109 Da)~\cite{Bionum}.
    \item[\textit{f}]  Based on rRNA sequences.  This value is slightly higher than that reported in \textit{E. coli} (324.3 Da)~\cite{Bionum}.
    \item[\textit{g}]  Calculated based on non-mitochondrial tRNA lengths, including introns.  See supplementary Excel file for individual tRNA lengths.
  \end{tablenotes}
\end{threeparttable}
\par \end{centering}
\label{table1}
\end{table*} 

\begin{table*}[h!] 
\noindent \begin{centering}
\begin{threeparttable}
\caption{\textbf{Reported ranges of biological parameters which are dependent on growth conditions}}
\begin{tabular}{|l|c|c|}
\hline 
 & Symbol & Value \tabularnewline
\hline
\hline
Growth rate~\cite{BoehlkeFriesen,Bonven79,WaldronLacroute75,Waldron77,Barkai} & $\mu$ & 0.086--0.46 hr$^{-1}$ \tabularnewline
\hline
Translation rate~\cite{Bionum}~\tnote{\textit{a}} & $k_{\text{ribo}}$ & 2.8--10.0 a.a./sec \tabularnewline
\hline
RNAP I transcription rate~\cite{FrenchRNAPIrates,Kos} & $k_{\text{RPI}}$  & 40--60 nt/sec \tabularnewline
\hline 
RNAP III transcription rate~\cite{FrenchRNAPIII} & $k_{\text{RPIII}}$ & 58--76 nt/sec \tabularnewline
\hline
Ribosomal activity (fraction of ribosomes which are active)~\cite{Bonven79,Barkai} & $f^{\text{active}}_{\text{ribo}}$ & 0.40--0.75 \tabularnewline
\hline
r-protein proteome fraction~\cite{Barkai}~\tnote{\textit{b}} & $\phi^{\text{r-prot}}_{\text{ribo}}$ & 0.14--0.32 \tabularnewline
\hline
Inferred fraction of active RNAPs I dedicated to 18S, 25S, 5.8S rRNA synthesis~\tnote{\textit{c}} \hspace{1mm} &  $\phi^{\text{m35S}}_{\text{RPI}}$ & $\sim \! 0.78$ \tabularnewline
\hline
Inferred fraction of active RNAPs III dedicated to 5S rRNA synthesis~\tnote{\textit{d}} &  $\phi^{\text{5S}}_{\text{RPIII}}$ & 0.11--0.14 \tabularnewline
\hline
\end{tabular}
  \begin{tablenotes}
    \item[\textit{a}]  There is a relatively wide range of reported translation rates, in part due to varying assumptions regarding the dependence of ribosomal activity vs. translation rate on growth rate.  We discuss the possibility of a constant translation rate of $k_{\text{ribo}} \approx 6.9$ a.a./sec; see subsection entitled ``Inferring the dependence of translation rate on growth rate in yeast'' in Section~\ref{VIII} for a discussion.
    \item[\textit{b}]  See Fig.~2A of Ref.~\cite{Barkai}.  
    \item[\textit{c}]  Because the only product of RNAP I is 35S, all active RNAPs I must be making 35S.  Flanking and spacer nucleotides account for $\approx\!22\%$ of the 35S pre-rRNA (see Fig.~\ref{rRNAprocessing}), and thus the remainder is $\phi^{\text{m35S}}_{\text{RPI}} \approx 78\%$.
    \item[\textit{d}]  RNAP III makes primarily 5S rRNA and tRNAs, aside from a few other small RNAs.  We estimate the fraction of active RNAPs III making 5S rRNA based on the number of tRNAs per ribosome in the cell (Table 4 of Ref.~\cite{WaldronLacroute75}) and the average tRNA length (Table~\ref{tab:fixednum}), as discussed at the end of Section~\ref{VIII}.
  \end{tablenotes}
  \label{tab:ranges}
\end{threeparttable}
\par \end{centering}
\end{table*} 

\clearpage

\end{widetext}


\clearpage

\begin{thebibliography}{9}


\bibitem{Klumpp2008} Klumpp, S. and Hwa, T., 2008. Growth-rate-dependent partitioning of RNA polymerases in bacteria. \textit{P. Natl. Acad. Sci.}, \textbf{105}(51), pp.20245-20250. 

\bibitem{ScottHwaReview} Scott, M. and Hwa, T., 2011. Bacterial growth laws and their applications. \textit{Curr. Opin. Biotech.}, \textbf{22}(4), pp.559-565.

\bibitem{KlumppPNAS} Klumpp, S., Scott, M., Pedersen, S. and Hwa, T., 2013. Molecular crowding limits translation and cell growth. Proceedings of the National Academy of Sciences, 110(42), pp.16754-16759.

\bibitem{Pugatch} Pugatch, R., 2015. Greedy scheduling of cellular self-replication leads to optimal doubling times with a log-Frechet distribution. \textit{P. Natl. Acad. Sci.}, \textbf{112}(8), pp.2611-2616. 

\bibitem{Alon1} Towbin, B.D., Korem, Y., Bren, A., Doron, S., Sorek, R. and Alon, U., 2017. Optimality and sub-optimality in a bacterial growth law. \textit{Nat. Commun.}, \textbf{8}, p.14123.

\bibitem{Alon2} Kohanim, Y.K., Levi, D., Jona, G., Towbin, B.D., Bren, A. and Alon, U., 2018. A bacterial growth law out of steady state. \textit{Cell Rep.}, \textbf{23}(10), pp.2891-2900.

\bibitem{Salman} Salman, H., Brenner, N., Tung, C.K., Elyahu, N., Stolovicki, E., Moore, L., Libchaber, A. and Braun, E., 2012. Universal protein fluctuations in populations of microorganisms. Physical review letters, 108(23), p.238105.

\bibitem{AmirNatComm} Lin, J. and Amir, A., 2018. Homeostasis of protein and mRNA concentrations in growing cells. \textit{Nat. Commun.}, \textbf{9}(1), p.4496.

\bibitem{AmirPRL} Amir, A., 2014. Cell size regulation in bacteria. \textit{Phys. Rev. Lett.}, \textbf{112}(20), p.208102.

\bibitem{Taheri-Araghi} Taheri-Araghi, S., Bradde, S., Sauls, J.T., Hill, N.S., Levin, P.A., Paulsson, J., Vergassola, M. and Jun, S., 2015. Cell-size control and homeostasis in bacteria. Current biology, 25(3), pp.385-391.

\bibitem{JunReview} Jun, S., Si, F., Pugatch, R. and Scott, M., 2018. Fundamental principles in bacterial physiology -- history, recent progress, and the future with focus on cell size control: a review. \textit{Rep. Prog. Phys.}, \textbf{81}(5), p.056601.

\bibitem{Woolford} Woolford, J.L. and Baserga, S.J., 2013. Ribosome biogenesis in the yeast Saccharomyces cerevisiae. Genetics, 195(3), pp.643-681.

\bibitem{Henras} Henras, A.K., Plisson‐Chastang, C., O'Donohue, M.F., Chakraborty, A. and Gleizes, P.E., 2015. An overview of pre‐ribosomal RNA processing in eukaryotes. Wiley Interdisciplinary Reviews: RNA, 6(2), pp.225-242.

\bibitem{Thomson} Thomson, E., Ferreira-Cerca, S. and Hurt, E., 2013. Eukaryotic ribosome biogenesis at a glance. Journal of Cell Science, 126, pp.4815-4821.

\bibitem{BenShem} Ben-Shem, A., Jenner, L., Yusupova, G. and Yusupov, M., 2010. Crystal structure of the eukaryotic ribosome. Science, 330(6008), pp.1203-1209.

\bibitem{Karbstein} Strunk, B.S. and Karbstein, K., 2009. Powering through ribosome assembly. Rna, 15(12), pp.2083-2104.

\bibitem{Dinman} Dinman, J.D., 2009. The eukaryotic ribosome: current status and challenges. Journal of Biological Chemistry, 284(18), pp.11761-11765.

\bibitem{Barkai} Metzl-Raz, E., Kafri, M., Yaakov, G., Soifer, I., Gurvich, Y. and Barkai, N., 2017. Principles of cellular resource allocation revealed by condition-dependent proteome profiling. Elife, 6, p.e28034.

\bibitem{Scott2010} Scott, M., Gunderson, C.W., Mateescu, E.M., Zhang, Z. and Hwa, T., 2010. Interdependence of cell growth and gene expression: origins and consequences. Science, 330(6007), pp.1099-1102.

\bibitem{Scott} Scott, M., Klumpp, S., Mateescu, E.M. and Hwa, T., 2014. Emergence of robust growth laws from optimal regulation of ribosome synthesis. \textit{Mol. Syst. Biol.}, \textbf{10}(8). 

\bibitem{Dai} Dai, X. and Zhu, M., 2020. Coupling of Ribosome Synthesis and Translational Capacity with Cell Growth. Trends in Biochemical Sciences.

\bibitem{Bionum} Milo, R., Jorgensen, P., Moran, U., Weber, G. and Springer, M., 2010. BioNumbers—the database of key numbers in molecular and cell biology. Nucleic acids research, 38(suppl 1), pp.D750-D753.

\bibitem{Dill} Dill, K.A., Ghosh, K. and Schmit, J.D., 2011. Physical limits of cells and proteomes. Proceedings of the National Academy of Sciences, 108(44), pp.17876-17882.

\bibitem{REP} Reuveni, S., Ehrenberg, M. and Paulsson, J., 2017. Ribosomes are optimized for autocatalytic production. Nature, 547(7663), pp.293-297.

\bibitem{KR} Kostinski, S. and Reuveni, S., 2020. Ribosome composition maximizes cellular growth rates in E. coli. Physical Review Letters, 125, 028103.

\bibitem{Speed-Limit}Klumpp, S., 2020. Speed Limit for Cell Growth. Physics, 13, p.108.

\bibitem{NomuraThoughts} Nomura, M., 1999. Regulation of ribosome biosynthesis in Escherichia coli and Saccharomyces cerevisiae: diversity and common principles. Journal of bacteriology, 181(22), pp.6857-6864. 

\bibitem{Warner} Warner, J.R., 1999. The economics of ribosome biosynthesis in yeast. Trends in biochemical sciences, 24(11), pp.437-440.

\bibitem{WaldronLacroute75} Waldron, C. and Lacroute, F., 1975. Effect of growth rate on the amounts of ribosomal and transfer ribonucleic acids in yeast. Journal of bacteriology, 122(3), pp.855-865.

\bibitem{MiloBook} Milo, R. and Phillips, R., 2015. Cell biology by the numbers. Garland Science.

\bibitem{ProteinDegradation} Martin-Perez, M. and Vill{\'e}n, J., 2017. Determinants and regulation of protein turnover in yeast. Cell systems, 5(3), pp.283-294.

\bibitem{SGD} Cherry, J.M., Hong, E.L., Amundsen, C., Balakrishnan, R., Binkley, G., Chan, E.T., Christie, K.R., Costanzo, M.C., Dwight, S.S., Engel, S.R., Fisk, D.G., Hirschman, J.E., Hitz, B.C., Karra, K., Krieger, C.J., Miyasato, S.R., Nash, R.S., Park, J., Skrzypek, M.S., Simison, M., Weng, S., and Wong, E.D., 2012. Saccharomyces Genome Database: the genomics resource of budding yeast. Nucleic acids research, 40(D1), pp.D700-D705.

\bibitem{KiefWarner} Kief, D.R. and Warner, J.R., 1981. Coordinate control of syntheses of ribosomal ribonucleic acid and ribosomal proteins during nutritional shift-up in Saccharomyces cerevisiae. Molecular and cellular biology, 1(11), pp.1007-1015.

\bibitem{BremerDennis} Bremer, H. and Dennis, P. 2008. Modulation of Chemical Composition and Other Parameters of the Cell at Different Exponential Growth Rates, EcoSal Plus 3(1), p.3. 

\bibitem{Boisvert} Boisvert, F.M., Ahmad, Y., Gierli{\'n}ski, M., Charri{\'e}re, F., Lamont, D., Scott, M., Barton, G. and Lamond, A.I., 2012. A quantitative spatial proteomics analysis of proteome turnover in human cells. Molecular \& Cellular Proteomics, 11(3).

\bibitem{Gawron} Gawron, D., Ndah, E., Gevaert, K. and Van Damme, P., 2016. Positional proteomics reveals differences in N‐terminal proteoform stability. Molecular systems biology, 12(2), p.858.

\bibitem{Waldron77} Waldron, C.,  Jund, R., and Lacroute, F., 1977. Evidence for a high proportion of inactive ribosomes in slow-growing yeast cells. Biochemical Journal, 168(3), pp.409-415.

\bibitem{Koch71} Koch, A.L., 1971. The adaptive responses of Escherichia coli to a feast and famine existence. In Advances in microbial physiology (Vol. 6, pp. 147-217). Academic Press.

\bibitem{BoehlkeFriesen} Boehlke, K.W. and Friesen, J.D., 1975. Cellular content of ribonucleic acid and protein in Saccharomyces cerevisiae as a function of exponential growth rate: calculation of the apparent peptide chain elongation rate. Journal of bacteriology, 121(2), pp.429-433.

\bibitem{Bonven79} Bonven, B. and Gull{\o}v, K., 1979. Peptide chain elongation rate and ribosomal activity in Saccharomyces cerevisiae as a function of the growth rate. Molecular and General Genetics MGG, 170(2), pp.225-230.

\bibitem{Piques} Piques, M., Schulze, W.X., H{\"o}hne, M., Usadel, B., Gibon, Y., Rohwer, J. and Stitt, M., 2009. Ribosome and transcript copy numbers, polysome occupancy and enzyme dynamics in Arabidopsis. Molecular systems biology, 5(1), p.314.

\bibitem{FrenchRNAPIrates} French, S.L., Osheim, Y.N., Cioci, F., Nomura, M. and Beyer, A.L., 2003. In exponentially growing Saccharomyces cerevisiae cells, rRNA synthesis is determined by the summed RNA polymerase I loading rate rather than by the number of active genes. Molecular and cellular biology, 23(5), pp.1558-1568.

\bibitem{Kos} Ko{\v s}, M. and Tollervey, D., 2010. Yeast pre-rRNA processing and modification occur cotranscriptionally. Molecular cell, 37(6), pp.809-820.

\bibitem{Sherman} Sherman, F., 2002. Getting started with yeast. In Methods in enzymology (Vol. 350, pp. 3-41). Academic Press.

\bibitem{Melnikov} Melnikov, S., Ben-Shem, A., De Loubresse, N.G., Jenner, L., Yusupova, G. and Yusupov, M., 2012. One core, two shells: bacterial and eukaryotic ribosomes. Nature structural \& molecular biology, 19(6), p.560.

\bibitem{FrenchRNAPIII} French, S.L., Osheim, Y.N., Schneider, D.A., Sikes, M.L., Fernandez, C.F., Copela, L.A., Misra, V.A., Nomura, M., Wolin, S.L. and Beyer, A.L., 2008. Visual analysis of the yeast 5S rRNA gene transcriptome: regulation and role of La protein. Molecular and cellular biology, 28(14), pp.4576-4587.

\bibitem{UnificationProt} Ho, B., Baryshnikova, A. and Brown, G.W., 2018. Unification of protein abundance datasets yields a quantitative Saccharomyces cerevisiae proteome. Cell systems, 6(2), pp.192-205.


\bibitem{Moll} Sauert, M., Temmel, H. and Moll, I., 2015. Heterogeneity of the translational machinery: Variations on a common theme. Biochimie, 114, pp.39-47.

\bibitem{EukaryaProteomeStable} Weiss, M., Schrimpf, S., Hengartner, M.O., Lercher, M.J. and von Mering, C., 2010. Shotgun proteomics data from multiple organisms reveals remarkable quantitative conservation of the eukaryotic core proteome. Proteomics, 10(6), pp.1297-1306.

\bibitem{Pelletier} Pelletier, J., Thomas, G. and Volarević, S., 2018. Ribosome biogenesis in cancer: new players and therapeutic avenues. Nature Reviews Cancer, 18(1), pp.51-63.

\bibitem{White} White, R.J., 2005. RNA polymerases I and III, growth control and cancer. Nature reviews Molecular cell biology, 6(1), pp.69-78.

\bibitem{CancerNucleoli} Quin, J.E., Devlin, J.R., Cameron, D., Hannan, K.M., Pearson, R.B. and Hannan, R.D., 2014. Targeting the nucleolus for cancer intervention. Biochimica et Biophysica Acta (BBA)-Molecular Basis of Disease, 1842(6), pp.802-816.

\end{thebibliography}
\end{document}